\documentclass[9pt,twocolumn,twoside]{osajnl}

\journal{ao}

\setboolean{shortarticle}{false}

\usepackage{upgreek}

\usepackage{acronym}
\acrodef{AO}[AO]{adaptive optics}
\acrodef{DM}[DM]{deformable mirror}
\acrodef{ELT}[ELTs]{extremely large telescopes}
\acrodef{ExAO}[ExAO]{extreme adaptive optics}
\acrodef{FOV}[FOV]{field of view}
\acrodef{FITS}[FITS]{Flexible Image Transport System}
\acrodef{FWHM}[FWHM]{full width at half maximum}
\acrodef{IFU}[IFU]{integral field unit}
\acrodef{IFS}[IFS]{integral field spectroscopy}
\acrodef{IRD}[IRD]{infrared doppler spectrograph}
\acrodef{LED}[LED]{light-emitting diode}
\acrodef{mas}[mas]{milli-arcseconds}
\acrodef{MCF}[MCF]{multi-core fiber}
\acrodef{MMF}[MMF]{multi-mode fiber}
\acrodef{MM}[MM]{multi-mode}
\acrodef{MLA}[MLA]{micro-lens array}
\acrodef{MFD}[MFD]{mode-field diameter}
\acrodef{NA}[NA]{numerical aperture}
\acrodef{NIR}[NIR]{near-infrared}
\acrodef{OAP}[OAP]{off-axis parabola}
\acrodef{PGMEA}[PGMEA]{propylene-glycol-methyl-ether-acetate}
\acrodef{PSF}[PSF]{point spread function}
\acrodef{PIAA}[PIAA]{phase-induced amplitude apodization}
\acrodef{POP}[POP]{physical-optics propagation}
\acrodef{QE}[QE]{quantum efficiency}
\acrodef{RHEA}[RHEA]{replicable high-resolution exoplanet and asteroseismology spectrograph}
\acrodef{RMS}[RMS]{root-mean-square}
\acrodef{RV}[RV]{radial velocity}
\acrodef{SCExAO}[SCExAO]{Subaru Coronagraphic Extreme Adaptive Optics}
\acrodef{SEM}[SEM]{scanning electron microscopy}
\acrodef{SMF}[SMF]{single-mode fiber}
\acrodef{SM}[SM]{single-mode}
\acrodef{SNR}[SNR]{signal to noise ratio}
\acrodef{SR}[SR]{strehl ratio}
\acrodef{VPH}[VPH]{volume phase holographic}
\acrodef{VSI}[VSI]{vertically-scanned white-light interferometry}
\acrodef{WFS}[WFS]{wavefront sensor}
\acrodef{3D-M3}[3D-M3]{3D-printed mono-mode multi-core fiber spectrograph}

\newcommand{\new}[1]{\textcolor{black}{#1}}

\title{3D-M3: High-spatial resolution spectroscopy
with extreme AO and 3D printed micro-lenslets}

\author[1,2,3,16,*]{Theodoros Anagnos}
\author[4,5]{Mareike Trappen}
\author[1,2]{Blaise C. Kuo Tiong}
\author[1,11]{Tobias Feger}
\author[7]{Stephanos Yerolatsitis}
\author[3]{Robert J. Harris}
\author[8]{Julien Lozi}
\author[13]{Nemanja Jovanovic}
\author[7]{Tim A. Birks}
\author[8]{Sébastien Vievard}
\author[8]{Olivier Guyon}
\author[7,12]{Itandehui Gris-S\'{a}nchez}
\author[9,10]{Sergio G. Leon-Saval}
\author[9,10]{Barnaby Norris}
\author[14,15]{Sebastiaan Y. Haffert}
\author[3]{Phillip Hottinger}
\author[4,5]{Matthias Blaicher}
\author[4,5]{Yilin Xu}
\author[9,10]{Christopher H. Betters}
\author[4,5,6]{Christian Koos}
\author[1,2]{David W. Coutts}
\author[1,2]{Christian Schwab}
\author[3]{Andreas Quirrenbach}

\affil[1]{Department of Physics and Astronomy, Macquarie University, NSW 2109, Australia}
\affil[2]{MQ Photonics Research Centre, Department of Physics and Astronomy, Macquarie University, NSW 2109, Australia}
\affil[3]{Landessternwarte, Zentrum f\"ur Astronomie der Universit\"at Heidelberg, K\"onigstuhl 12, 69117 Heidelberg, Germany}
\affil[4]{Institute of Microstructure Technology (IMT), Karlsruhe Institute of Technology (KIT), Hermann-von-Helmholtz-Platz 1, 76344 Eggenstein-Leopoldshafen, Germany}
\affil[5]{Institute of Photonics and Quantum Electronics (IPQ), Karlsruhe Institute of Technology (KIT), Engesserstr. 5, 76131 Karlsruhe}
\affil[6]{Vanguard Photonics GmbH, Hermann-von-Helmholtz-Platz 1,76344 Eggenstein-Leopoldshafen, 76227 Karlsruhe}
\affil[7]{Department of Physics, University of Bath, Claverton Down, Bath, BA2 7AY, UK}
\affil[8]{National Institutes of Natural Sciences, Subaru Telescope, National Astronomical Observatory of Japan, Hilo, Hawaii, United States}
\affil[9]{University of Sydney, Sydney Astrophotonics Instrumentation Laboratory, School of Physics, Camperdown, Australia}
\affil[10]{University of Sydney, Sydney Institute for Astronomy, School of Physics, Camperdown, Australia}
\affil[11]{Redback Systems Pty Ltd, Sydney, Australia}
\affil[12]{currently at ITEAM Research Institute, Universitat Polit\`{e}cnica de Val\`{e}ncia, Camino de Vera, 46022 Valencia, Spain}
\affil[13]{California Institute of Technology, 1200 E. California Blvd., Pasadena CA, 91125, USA}
\affil[14]{Leiden Observatory, Leiden University, PO Box 9513, Niels Bohrweg 2, 2300 RA Leiden, The Netherlands}
\affil[15]{Steward Observatory, University of Arizona, 933 North Cherry Avenue, Tucson, Arizona}
\affil[16]{Max-Planck-Institut f\"ur Astronomie, K\"onigstuhl 17, 69117, Heidelberg, Germany}
\affil[*]{anagnos.theodoros@gmail.com}
\affil[*]{\copyright 2021 Optical Society of America]. One print or electronic copy may be made for personal use only. Systematic reproduction and distribution, duplication of any material in this paper for a fee or for commercial purposes, or modifications of the content of this paper are prohibited.}

\begin{abstract}
By combining \acl{IFS} with \acl{ExAO} we are now able
to resolve objects close to the diffraction-limit of
large telescopes, exploring new science cases. We
introduce an \acl{IFU} designed to couple light with
a minimal platescale from the \acs{SCExAO} facility 
at \acs{NIR} wavelengths to a \acl{SM} spectrograph.
The \acl{IFU} has a 3D-printed \acl{MLA} on top of a
custom \acl{SM} \acl{MCF}, to optimize the coupling
of light into the fiber cores. We demonstrate the 
potential of the instrument via initial results from 
the first on-sky runs at the 8.2 m Subaru Telescope
with a spectrograph using off-the-shelf optics, 
allowing for rapid development with low cost.
\end{abstract}

\setboolean{displaycopyright}{true}

\begin{document}

\maketitle

\section{Introduction}
A plethora of intrinsic and extrinsic information can
be collected through astronomical spectroscopy, such 
as distance, motion properties, chemical characteristics
as well as the existence of nearby celestial bodies
\cite{Massey:2013}. New techniques are rapidly 
advancing to extract more information from the 
spectral data, which translates to discoveries
about celestial phenomena.

Almost half a century ago, the first fiber-fed
instruments began supplementing existing astronomical
spectrographs (e.g., Ref.~
\cite{Gray:1982,Hubbard:1979,Powell:1984}).
With the help of optical fibers, it became possible
to guide starlight from the telescope to remotely
located instruments. This relaxed the mechanical
constraints due to varying gravity vectors affecting
the spectrograph, and the measurement precision
increased as the instrument could be stabilized and 
controlled far better in a stable environment as 
opposed to that in the telescope dome. The fibers in
these instruments are typically large in diameter 
\acp{MMF}, in order to couple as much light as possible
from a seeing-limited image \cite{Hill:1980,Hill:1988}.

A contemporary development of fiber spectrographs
was the \acf{IFU}. A fibered \ac{IFU} makes
use of several fibers to provide separate spatial
samples, to investigate small regions of the image
plane (e.g. adjacent or extended objects)
\cite{Allington-Smith:2006}. The spectra from different
locations of the image plane are positioned 
on the same detector, but physically displaced to avoid
cross contamination. The use of fibers for \ac{IFS}
provides spatial information of the target object, 
as well as improved flexibility to reformat the obtained
spectra to fit the limited detector space. 

Using multiple \acp{MMF} with relatively large core
diameters of greater than $\sim$50\,$\upmu$m to form
an \ac{IFU} that couples seeing-limited starlight,
is today a relatively standard practice (e.g., Ref.
\cite{Ge:1998,Croom:2012,Bryant:2016,Tozzi:2018}),
which has enabled many discoveries. However, these
instruments are large, due to conservation of \'{e}tendue,
which means they are costly and difficult to stabilize.
The ability to reduce the size would open up new 
avenues and reduce cost. One solution which has been
proposed for several types of instruments (e.g., Ref.
\cite{Bland-Hawthorn:2010,Leon-Saval:2012,Betters:2013, Betters:2014,Tamura:2012,Crepp:2016,Rains:2016}),
is to use fibers with smaller core diameter, or \acp{SMF}.
The main roadblock is that coupling starlight into
\acp{SMF} from telescopes with a 4\,m or larger 
primary diameter with a seeing-limited output\new{, scale 
that is larger than the Fried parameter $r_{0}$,} is 
extremely inefficient.

Various efforts have been undertaken to increase the
coupling efficiency to \acp{SMF}, with great success 
in recent years (e.g., Ref.~
\cite{Coude:1992,Coude:2000,Ghasempour:2012,Harris:2015,Bechter:2016,Crepp:2016,Cvetojevic:2017,Jovanovic:2017,Anagnos:2018,Bechter:2020}).
This is in large part due to \ac{AO} systems, which
reduce the \ac{PSF} size and wavefront deformation 
to acceptable levels. In particular, \ac{ExAO} systems
have demonstrated over $90\%$ Strehl ratio correction 
(restoring the \ac{PSF} close to the diffraction-limit)
in the H-band (1.5-1.8\,$\upmu$m) (e.g., Ref.
\cite{Dekany:2013,Agapito:2014,Macintosh:2014,Jovanovic:2015,Vigan:2016,Guyon:2020}), allowing close to
optimum coupling. \ac{ExAO} systems are also able to
maintain a constant \acs{PSF} position, in better than
median seeing on-sky conditions, which is essential 
for efficient coupling \cite{Ruilier:1998}.

Using \acp{SMF} also enables astronomers to take advantage
of the diffraction-limited imaging provided by \ac{AO}.
Together these can open up science cases not available
to conventional instruments. Examples include spatially
resolving the surface of barely resolved stellar photospheres
\new{as well as their chemical composition of their close molecular
layer known as MOLsphere}. The \ac{SCExAO} system at the Subaru Telescope is an
ideal platform for combining such approaches \new{\cite{Jovanovic:2015}}.
The list of stars that could be imaged includes (in an angular
diameter size order), Betelgeuse ($\alpha$ Orionis) 
$\sim$50\,\ac{mas}, Mira ($o$ Ceti) $\sim$50\,\ac{mas},
W Hya $\sim$42\,\ac{mas}, Antares ($\alpha$ Scorpii)
$\sim$40\,\ac{mas}, $\alpha$ Her $\sim$36\,\ac{mas},
R Leo $\sim$31\,\ac{mas}, Arcturus ($\alpha$ Bootis)
$\sim$21\,\ac{mas}, $\mu$ Cep $\sim$21\,\ac{mas},
$\gamma$ Her $\sim$21\,\ac{mas} and Aldebaran ($\alpha$
Tau) $\sim$19\,\ac{mas}.
These are mostly giants with complex mass loss through
a cool ($\sim$2000 K) molecular wind ($\sim$10 km 
$\mathrm{s^{-1}}$) and/or a hot corona ($\sim$10,000 K)
\cite{Tsuji:2008}. A spectrograph with a resolving power
of several tens of thousand is enough to distinguish the
differential spectral velocities of the surface that are
of order a few km $\mathrm{s^{-1}}$ and to probe the
atmospheric structure by measuring the wavelength-dependence
of the stellar diameter \cite{Quirrenbach:1993}. A single
observation will provide useful information regarding
the photosphere of the star in high resolution, while
a sequence of observations can inform on the differential
rotation of the surface features, probing the magnetic 
field properties and the mass loss of the target (e.g., Ref.
\cite{Lobel:1999,Lobel:2004,Ohnaka:2009,Khouri:2018,Ohnaka:2018,Wood:2019}).
Further science applications involve the study of Mira
variables where the mass loss kinematics are of special
importance as well as their dust and wind properties
(e.g., Ref.~\cite{Norris:2012}), and investigation of the 
orbital parameters of sub-arcsecond companions such as
the unevenly bright yet well resolved Mira AB system,
which is the closest wind accretion binary (e.g., Ref.
\cite{Karovska:2011,Ramstedt:2014,Vlemmings:2015}).
Additionally, with a high resolution spectrograph
the stellar chromospheres can be studied by
observing chromospherically sensitive lines, such as
the Paschen lines in the infrared (IR) regime and the
He\,I IR triplet lines at 1083\,nm, which are valuable 
indicators of chromospheric activity \cite{Fuhrmeister:2019}.
Lastly, stabilizing the environment of the instrument
will enable precise \ac{RV} measurements.

Whilst \ac{SM} \acp{IFU} enable high-spatial 
resolution spectroscopy and open these new science cases,
they suffer from an issue with
fill fraction. Due to the need for a cladding region
around the central core, fiber \acp{IFU} can only sample
a certain percentage of the light at the focal plane.
Originally the percentage of sampled light was of the 
order of 60-65\% for \ac{MMF}-\acp{IFU}, though recent
developments have increased this to 73\% \cite{Croom:2012}.
To increase the coupling efficiency, microlenses can be
attached to the \ac{IFU}, increasing the fill fraction 
(e.g., Ref.~\cite{Courtes:1982}).
However, aligning these arrays and fiber bundles is
difficult as the alignment tolerances are of the order 
of a few microns.

To reduce these problems, an alternative and 
more versatile type of fiber that can act as an \ac{IFU},
i.e. the \ac{MCF} could be utilized. This is a combination
of at least two fiber cores forming an individual fiber
using a common cladding.
The cores of a \ac{MCF} can sample light across the image
plane making it ideal for constructing an \ac{IFU}. By 3D
printing lens structures on top of its cores using
lithography techniques, we save time and avoid alignment issues,
dramatically improving the free-space coupling of light
into the fiber and relaxing the requirement for precise
alignment of the input (e.g., Ref.~
\cite{Gissibl:2016,Dietrich:2016,Dietrich:2017,Dietrich:2018,Hottinger:2018}).

Currently, the \ac{RHEA} at Subaru 
\cite{Rains:2016,Rains:2018} is \new{one of the few \ac{IFU}
configured instruments using \acp{SMF}
\cite{Crepp:2016,Mawet:2016,Haffert:2020}}. It uses the
visual arm for wavelengths below 900\,nm, of the 
\ac{SCExAO} system, but suffers from low coupling
efficiency and the presence of modal noise. The low
coupling efficiency is likely due to mis-alignments
in the commercial off-the-shelf \ac{MLA} and anamorphic
prisms to couple the light into the cores of the
bundle of the fibers (stacked and glued together),
which are sensitive to alignment, illustrating the
difficulties of applying conventional approaches
to \ac{SM} technology. \new{Following a similar approach
as detailed below, the \ac{IFU} of \ac{RHEA} was
upgraded recently with 3D-printed \acp{MLA}
on a \ac{MCF} in the visual band \cite{Anagnos:2020}.}

In this paper, we show the first demonstration
of a custom \ac{MCF} with in-situ 3D printed \ac{MLA}
on top of its 19 cores using two-photon polymerization 
lithography, in combination with a high resolving
power compact \ac{SM} spectrograph optimized for the
900\,-\,1100\,nm wavelength range,  further on referred
to as \acf{3D-M3}. This spectrograph takes design cues
from \ac{RHEA}@Subaru \cite{Rains:2016} for a wavelength
range in the \ac{NIR} and the multicore fiber Photonic
TIGER \'Echelle spectrograph 
\cite{Leon-Saval:2012,Betters:2014}.
The instrument is designed for operation in the
\ac{NIR} arm of the \ac{SCExAO} facility, making
\ac{3D-M3} the first instrument to take advantage
of the \ac{PIAA} optics in that path. \ac{PIAA} 
are custom optics \cite{Lozi:2009} that induce a 
softer apodization on the collimated beam, forming
a near-Gaussian profile \ac{PSF} output in the image
plane \cite{Jovanovic:2017}. This makes the coupling
of starlight more efficient as the \ac{PIAA}
beam intensity profile is close to the mode-field
profile of an \ac{SMF}. The setup was tested on-sky
using \ac{SCExAO} at the 8.2\,m Subaru Telescope;
it covers an instantaneous 54\,\ac{mas} \ac{FOV}. 
The custom \ac{MCF} achieves high throughput coupling
by using the 3D printed \ac{MLA} on top of each of 
its cores, which provide a very high fill factor of 
the fiber surface and are extremely efficient for 
off-axis field injection while acting as an \ac{IFU}.
Combined with the \ac{SM} spectrometer and \ac{SCExAO},
the instrument shows potential towards a precise
high-resolution \ac{IFS} with astrophysical sources 
due to its inherent \ac{SMF} stability and absence of
modal noise as the \ac{MCF} is designed to exhibit 
\ac{SM} behavior in the wavelength regime of operation.

In Section \ref{sec:methods} we discuss the conceptual
design and parameters, followed by a description of
the experimental setup to characterize performance. 
Then in Section \ref{sec:results} we present the
laboratory and on-sky results. Finally, we summarize 
in Section \ref{sec:discussion}.

\section{Methods}
\label{sec:methods}
To efficiently inject light from an 8-m class telescope
into the cores of a \ac{MCF} and a diffraction-limited
spectrograph, the following elements are required: a high
performance \ac{AO} system, an \ac{IFU} with micro-lenses
to dissect the focal plane image and direct light to the
individual cores efficiently, the \ac{MCF} itself,
and the spectrograph, which must be able to handle
a multi-core input In this section these elements will
be discussed in detail.

\subsection{Instrument architecture}
\label{sec:sim-arch}
To simulate the instrument, we compile component
specifications from the resulting beam output of
the telescope down to the detector. Starting with 
the 8-m Subaru Telescope, starlight continues through
the complex set of the AO systems at Subaru, namely
AO188 and \ac{SCExAO}, where the intensity distribution of the
resulting output beam is close to a Gaussian profile. 

\begin{figure}
\centering
\includegraphics[width=\columnwidth]{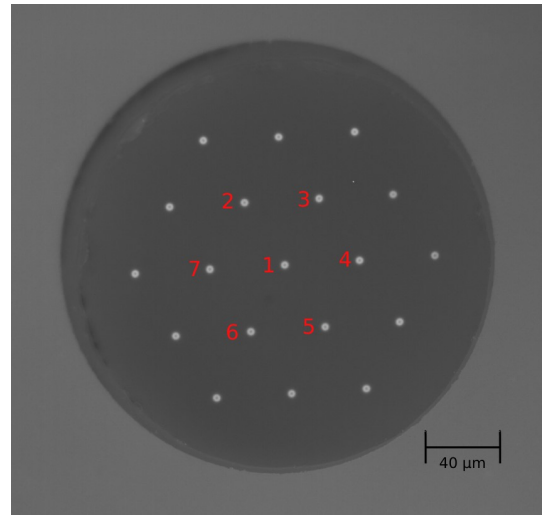}
\caption{Microscope image of the polished \acl{MCF} 
in its ferrule. The cores are visible in a hexagonal
configuration with a pitch of 40\,$\upmu$m. The fiber has
a 5.3\,$\upmu$m \acl{MFD} at 980\,nm ($\mathrm{1/e^{2}}$)
and a cladding diameter of $\sim$ 220\,$\upmu$m. Here,
the cores are back-illuminated with a white light source.
Numbers in red are used for core referencing below.}
\label{fig:mcf-x-sect}
\end{figure}

To efficiently couple light into the \ac{SM}
cores of the \ac{MCF}, a custom fiber injection unit
is required. This system has to be carefully optimized
to match the parameters of the \ac{MCF}, which has
a 5.3\,$\upmu$m \ac{MFD} at 980\,nm ($\mathrm{1/e^{2}}$)
with a \ac{SM} cut-off at $\sim$800\,nm \new{with an upper
working wavelength limit of 1400\,nm }(a microscope
image of the cross-section of the fiber is presented in 
Fig.~\ref{fig:mcf-x-sect}). The core-to-core spacing 
(pitch) of the \ac{MCF} is 40\,$\upmu$m. This is 
important to make sure that the cross-coupling of 
light between the cores is negligible, as they are 
physically sufficiently separated with a ratio of 
pitch/\ac{MFD} of 7.5:1\new{, resulting in a theoretical 
cross-coupling less than few percent}. To maximize the 
amount of light coupled to the fiber cores, micro-optics 
are 3D printed on top of the cores using a custom 
lithography system described below. To demonstrate the 
potential of such an injection unit system, a compact
spectrograph was built with off-the-shelf optics detailed 
in the following sections.

To spatially resolve a stellar surface without compromising 
the throughput, the selected plate scale was $0.45''$/mm.
That means that each fiber core subtends 18\,\ac{mas}
on the sky.
As a reference, the Subaru Telescope with the \ac{SCExAO}
facility is able to spatially resolve the surface of 
stars with an angular diameter of $\sim$50\,\ac{mas},
which is the apparent size of a handful of giant stars,
such as, for example, Betelgeuse ($\alpha$ Ori),
or Mira ($o$ Ceti). The required focal ratio to achieve
the $0.45''$/mm plate scale is $f/57.5$. \new{The diffraction 
limit of the Subaru Telescope at 980\,nm is 25\,\ac{mas}.
The plate scale of $0.45''$/mm was selected for a sub-diffraction
limit sampling in order to seek for the sweet spot between
spatially resolve the stellar surface and perform decently
in terms of throughput.}

\subsection{The SCExAO infrastructure}
\label{sec:scexao-infr}

\begin{figure*}
\centering
\includegraphics[width=0.8\textwidth]{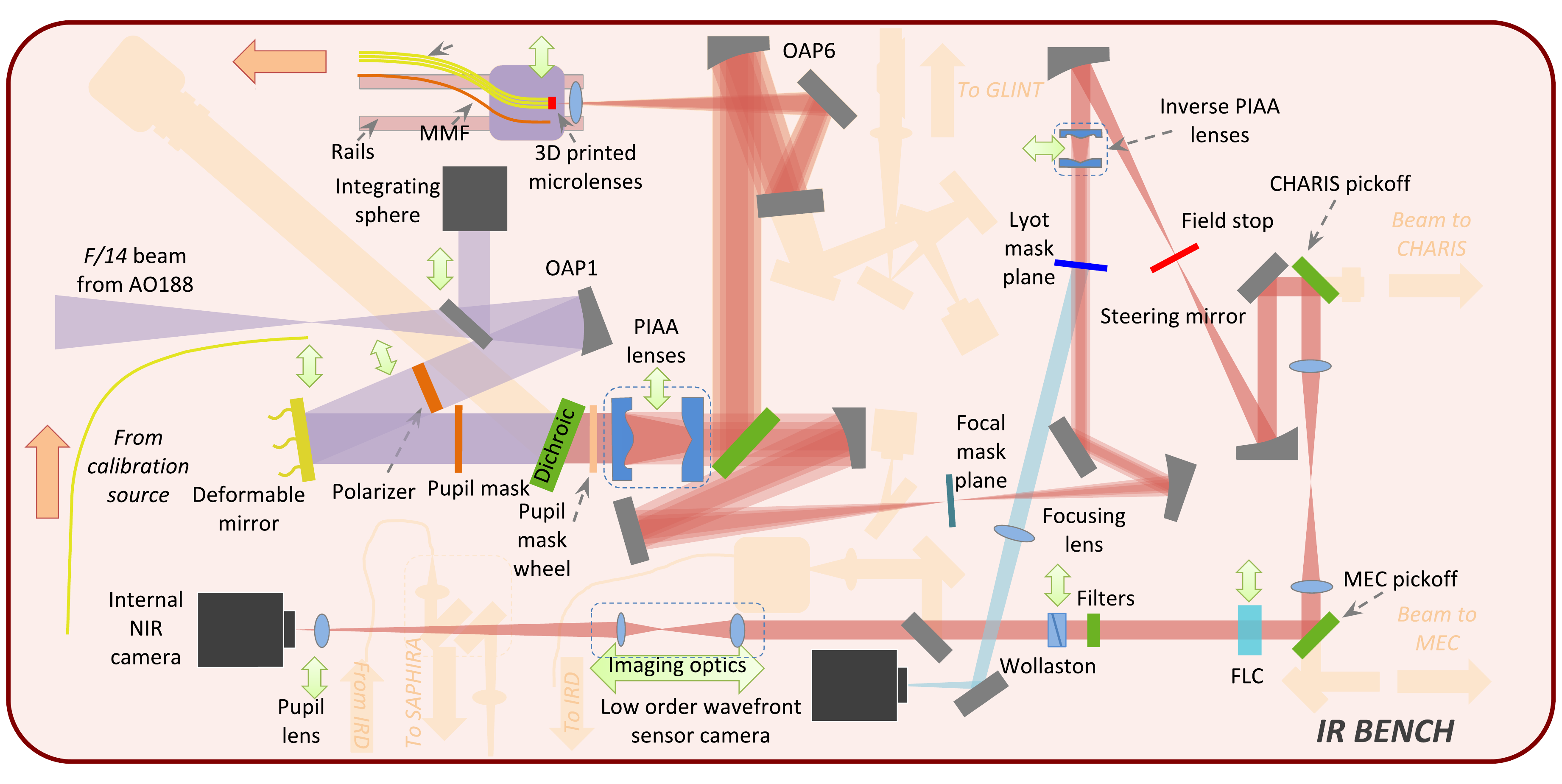}
\caption{Schematic illustration of the \ac{NIR}
bench of the \ac{SCExAO} facility. The 3D-printed 
fiber is shown at top left. The beam from the AO188
to the \ac{IFU} is represented with purple and red color.}
\label{fig:bench-scexao}
\end{figure*}

\ac{SCExAO} is installed at the \ac{NIR} 
Nasmyth focus of the 8.2\,m Subaru Telescope. Detailed
information and schematic layouts are provided
in \cite{Jovanovic:2015}. Initially the starlight 
from the 8.2\,m Subaru Telescope enters the AO188
facility, which offers 30-40\% Strehl \ac{PSF} 
correction in the H-band under moderate atmospheric
seeing conditions 
\cite{Hayano:2008,Hayano:2010,Minowa:2010}. After
a series of components in the optical train for
manipulation of the beam, the light undergoes
further wavefront corrections of higher-order
spatial and temporal modes caused by the atmosphere.
Next, it is filtered using a dichroic filter and 
split in two paths, the visible channel ($<$900\,nm)
and the \ac{NIR} ($>$900\,nm) channel. However, 
the wavelength response of the dichroic filter does
not have a sharp cut-off profile, and there is a
slight overlap of wavelengths between both channels.
After that, the starlight is focused using a 
gold-coated \ac{OAP} with $f$ = 519\,mm onto the 
3D-printed \ac{MLA} surface. The 3D-printed end of 
the \ac{MCF} was installed on the \ac{NIR} bench 
of \ac{SCExAO} (see Fig.~\ref{fig:bench-scexao}).

\subsection{Simulations}
\label{sec:sims}
To simulate the output beam, the profile of the 
Subaru Telescope and the key parameters of \ac{SCExAO}
from the literature (e.g. Ref.
\cite{Jovanovic:2015,Jovanovic:2017}) were used.

The \ac{POP} module of the commercial
\texttt{Zemax} \cite{zemax} ray tracing software was
used to simulate the output beam profile of the Subaru
Telescope, including the effect of the \ac{PIAA} optics
\cite{Lozi:2009} installed in the \ac{NIR} optical
train of \ac{SCExAO}, where the fiber is positioned.
This output was used as an input for 
optimizing the performance of the \ac{MLA} structure
that forms the \ac{IFU}.

To optimize the coupling efficiency of the \ac{SCExAO} 
beam output at its focal plane, the physical-optics
propagation module was used again to model the surface
shape of the \ac{MLA} structure. Even polynomials with
coefficients up to eighth order were used to model the
aspheric lens surfaces of the \ac{MLA}, forming a 
hexagonal aperture at the top end of the lenslets where
they are merged together \new{leaving no space in between}. 
Since the 3D printing method is able to achieve surface
roughness with an \ac{RMS} of 37\,nm \cite{Dietrich:2018},
the roughness of the structure is expected to be better
than $\lambda/20$ at the wavelength of operation. \new{This 
resulted to a total coupling efficiency of 48.8\%
(coupling efficiency 56.4\%, transmission through the 
\ac{MLA} 86.4\%, total coupling 48.8\%).}

\subsection{Micro-optic manufacturing process}
\label{sec:man-proc}

The structure of the \ac{MLA} \new{as detailed in 
Section~2.\ref{sec:sims} }was directly printed in
a single block using two-photon lithography into the
commercially available negative-tone photoresist IP-Dip
\cite{IP-Dip}. The \ac{MLA} was printed on the cleaved
facet of the \ac{MCF}, which had been manually glued into
an FC-PC connector and then polished to achieve a flat
surface to enable straight-forward printing on the front
face of the fiber. A lithography machine built in house
\new{\cite{Dietrich:2016,Dietrich:2017,Dietrich:2018,Blaicher:2020}}
was used to generate the lenses. The system is equipped
with a 780\,nm femtosecond laser \cite{Laser} and
a 40$\times$ Zeiss \new{oil immersed} objective lens with \ac{NA}\,=\,1.4.
For high-precision alignment and writing with high
shape fidelity, machine control software developed
in house was used. 

\begin{figure}
\centering
\includegraphics[width=\columnwidth]{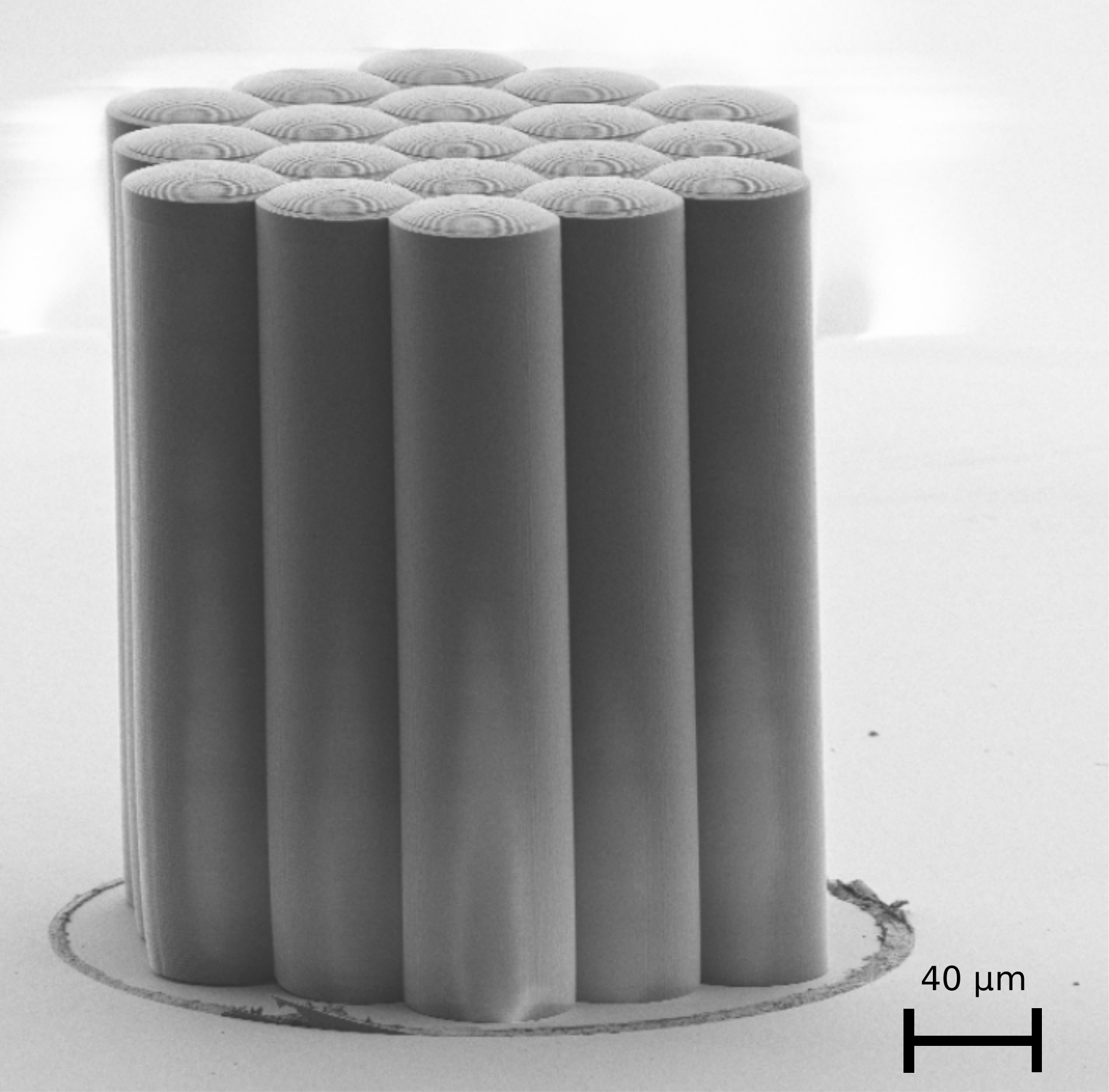}
\caption{Scanning electron microscope image of the 3D 
printed \acl{MLA} structure on top of the \acl{MCF} ferrule.
All 19 cores were 3D printed, though only 7 were used 
for the on-sky run.}
\label{fig:mla-sem}
\end{figure}

The fiber was back-illuminated with a red \acs{LED}, 
which was used together with machine vision to detect 
the 19 cores of the \ac{MCF} and align the individual
lenses to each core. To compensate for any
slight location and pitch variation of the individual
cores of the \ac{MCF}, the full 3D-model is generated
only after core detection. Prior to printing, the 
individual models of the lenses are merged by a
Boolean operation to avoid unnecessary double
illumination. Automated detection of the fiber 
end-face tilt is employed and the structures are
corrected accordingly. The writing distances between
subsequent lines and layers, i.e., both hatching and
slicing distance, were set to 100\,nm. The fabricated
structure is developed in \ac{PGMEA}, flushed with
isopropanol, and subsequently blow dried.
In the next stage, \ac{SEM} and \ac{VSI} images of
the structure were acquired to check the quality of 
the manufacturing process (see Fig.~\ref{fig:mla-sem}).

\subsection{Fiber injection unit}
\begin{figure}
\centering\centering
\includegraphics[width=\columnwidth]{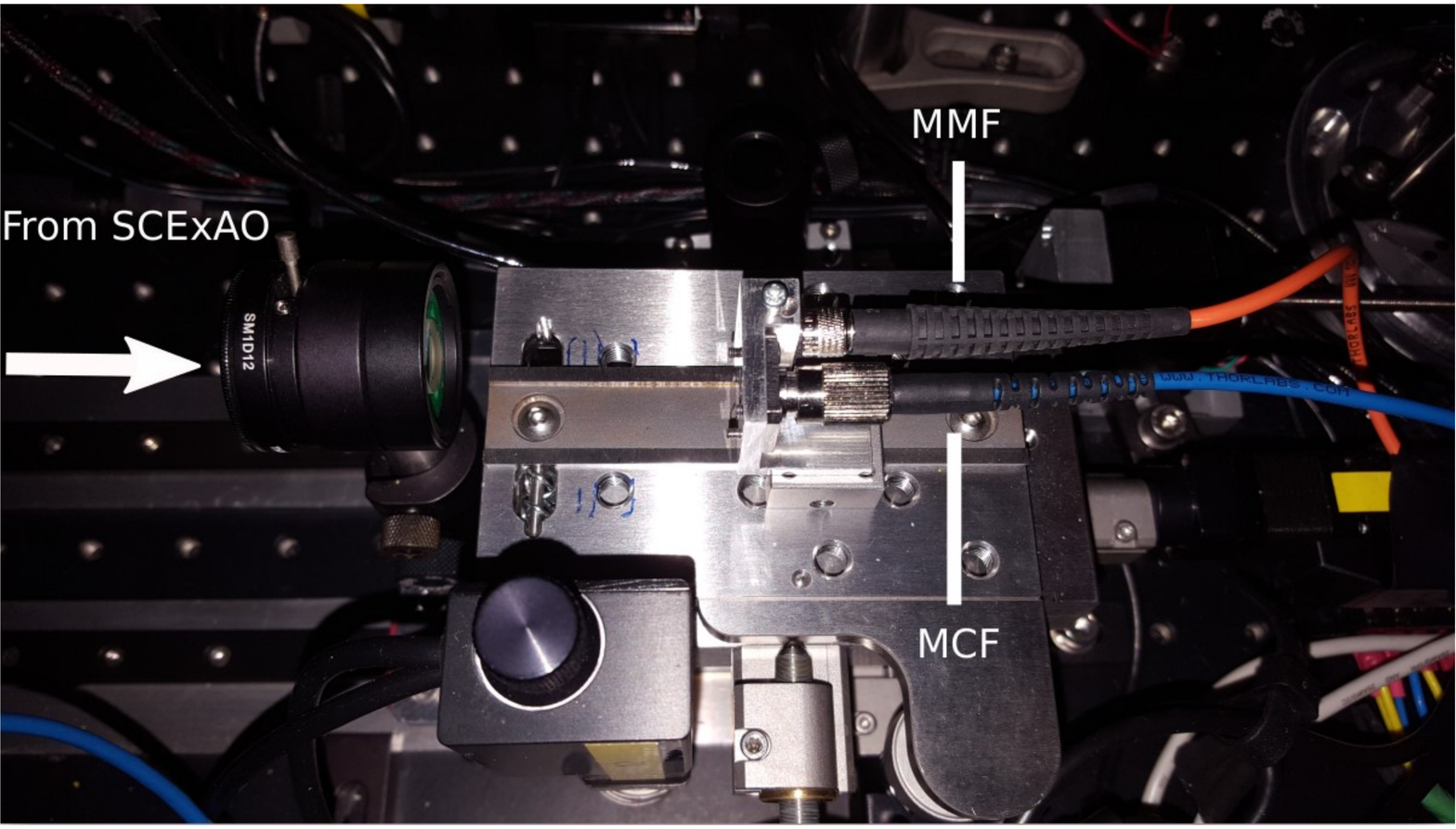}
\caption{Illustration of the fiber-injection  
opto-mechanic setup on the \ac{SCExAO} \ac{NIR} bench.
The \ac{MCF} (blue tubing) with the 3D-printed \ac{MLA}
is attached to a fiber socket and inserted in the beam 
path. An \ac{MMF} (orange tubing) can be inserted in 
the beam for calibration of the system.}
\label{fig:inj-scexao}
\end{figure}

In order to achieve precise repeatable positioning of 
the injection fiber across the \ac{PSF} of the target,
a 5-axis remotely controlled stage was used. The 5-axis
stage (Newport, M-562-XYZ \& 562F-TILT) was equipped 
with computer controlled stepper motors (Zaber, T-NA08A25),
which allowed a minimum step translation of $\sim$50\,nm
and $<$ 1\,$\upmu$m of unidirectional repeatability 
(see Fig.~\ref{fig:inj-scexao}). To align the fiber
for tip-tilt, adjustments were performed manually at
the L-bracket on which the fibers were mounted. The 
alignment of 
the fiber to the optical axis was achieved by adjusting
the Gaussian illumination at the pupil to be centered 
while the fiber was back-illuminated with a HeNe laser.
Immediately in front of the L-bracket, a $-$48\,mm focal 
length plano-concave lens (Edmunds\,\#67-995) was
installed to re-adjust the injected $f/\#$ to match that
of the fiber. This was accomplished by using rails to
translate the plano-concave lens and the 5-axis
mount together and independently of each other. Using
this arrangement a range of $f/\#$ from 28 to
60 could be achieved.

The L-bracket was designed to support 2 fibers,
the \ac{MCF} and another fiber with SMA connector.
On the SMA slot, a \ac{MMF} was installed 
(365\,$\upmu$m core, NA\,=\,0.22, Thorlabs - FG365LEC)
to calibrate the throughput measurements as the
5-axis stage moves laterally (sideways).

In order to achieve $f/\#$ of 57.5 (see Section~2.\ref{sec:sim-arch}),
the spacing between the
focusing \ac{OAP} (f\,=\,519\,mm), the concave
lens and the fiber was adjusted. \texttt{Zemax}
simulations served as a reference starting point,
and then further optimized using a detector imaging
the near field output of the \ac{MCF} to confirm
the required $f/\#$.

\subsection{Throughput measurement setup}
\label{sec:throughput}
\begin{figure}
\centering
\includegraphics[width=\columnwidth]{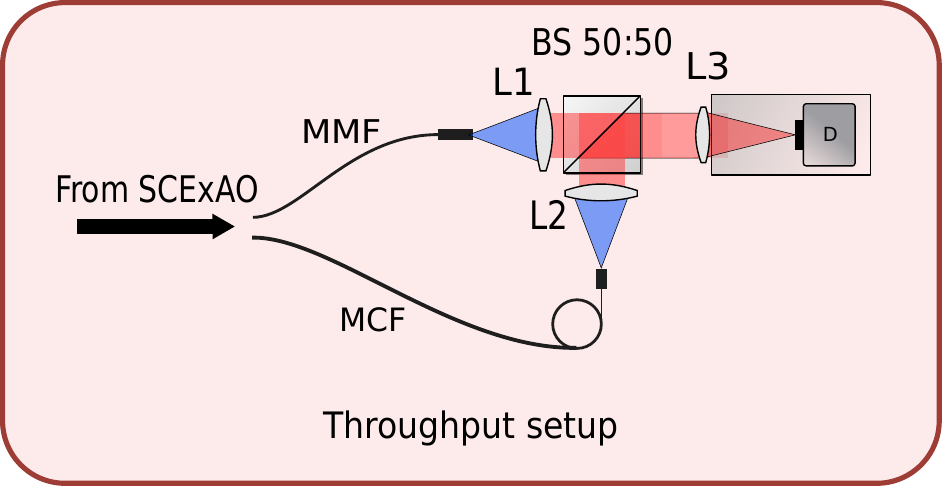}
\caption{Throughput test setup for measuring the 
efficiency of the 3D printed \acl{MLA}. Starlight from 
\ac{SCExAO} is sampled by one of the fibers at any time.
The calibration of the throughput is achieved using the
\acl{MMF}, lenses  (L1-L2-L3) for collimation and 
focusing of the beam, beamsplitter (BS) and CMOS detector 
(D) for sampling the light output.}
\label{fig:th_setup}
\end{figure}

To measure the total throughput performance of 
the custom \ac{IFU}, a throughput
test experiment was constructed. A schematic illustration
of this setup is presented in Fig.~\ref{fig:th_setup}.
The light from each of the fibers was collimated using
a set of microscope lenses; L1 - Thorlabs RMS20X-PF for
the \ac{MMF} and L2 - Thorlabs RMS4X-PF for the \ac{MCF}.
Both beams were directed towards a 50:50 non-polarizing
beamsplitter (BS) (Thorlabs CM1-BS014). After the BS, a
50\,mm achromatic lens (L3\,-\,Thorlabs\,AC254-050-B-ML) was
used to re-focus the light on the detector 
(D\,-\,ASI-120MM-S mono). To measure the total throughput
(including the coupling losses) of the \ac{MCF}, the
throughput of the \ac{MMF} was used to calibrate the
absolute flux. This was achieved by translating the 
5-axis stage at the telescope focus to couple
it to each core of the \ac{MCF} input, and then
normalizing each collected signal by the flux measured
in the \ac{MMF}.

\subsection{Spectroscopic setup}
\label{sec:spec}

To show the potential of using a 3D printed \ac{MLA} in
conjunction with an \ac{MCF}, a spectrograph similar to 
the \ac{RHEA} at Subaru design \cite{Feger:2014,Rains:2016}
and the MCF Photonic TIGER \'Echelle spectrograph
\cite{Leon-Saval:2012,Betters:2014} was built. The 
benchtop instrument is a compact diffraction-limited 
\'echelle spectrograph using only off-the-shelf components.

\begin{figure}
\centering
\includegraphics[width=\columnwidth]{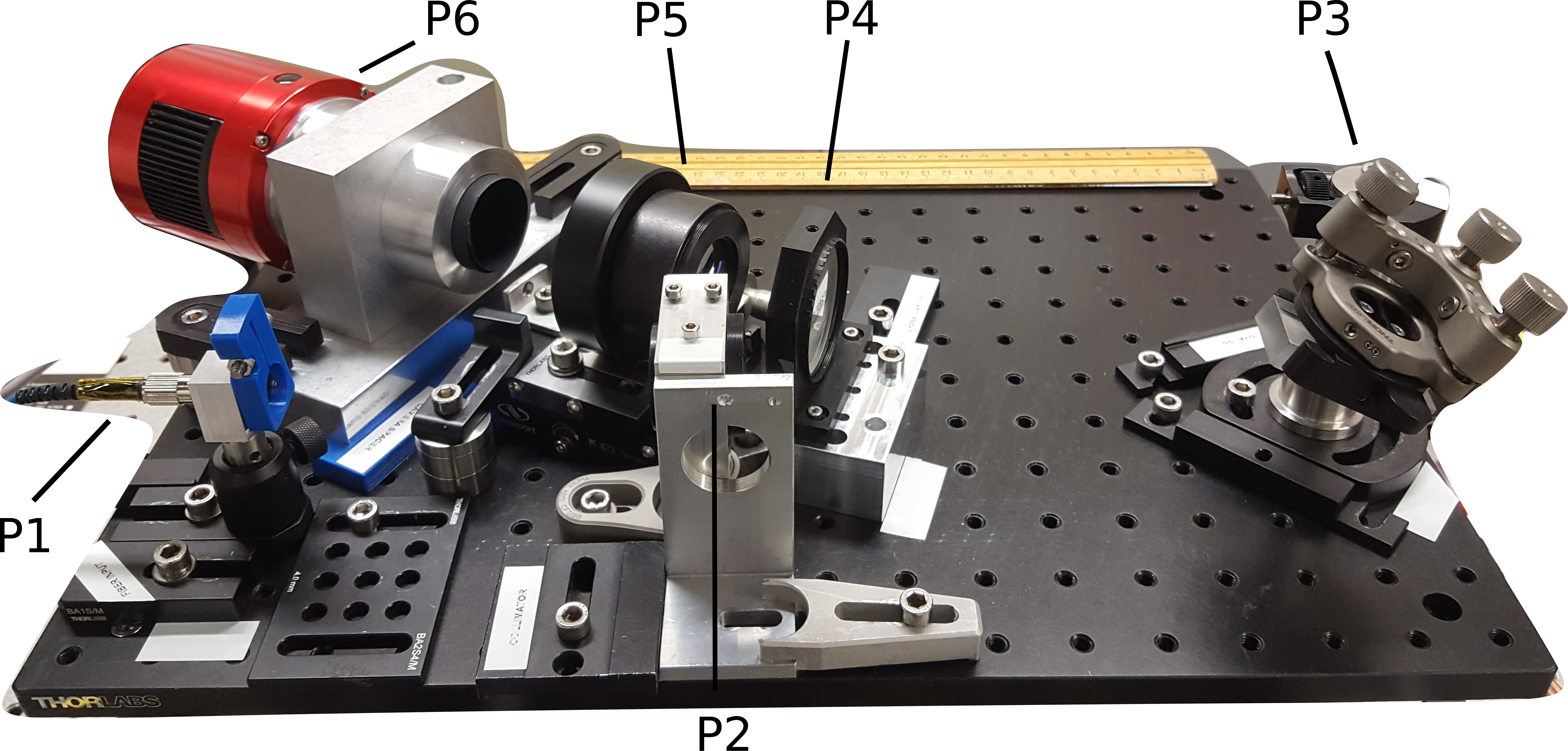}
\caption{Illustration of the spectroscopic setup of 
the \ac{3D-M3} using only off-the-shelf components.
The footprint of the breadboard is $300\times600$\,mm.
The light injected from the \ac{MCF} (P1) is collimated
by a combination of optical lenses (P2) before it is
chromatically dispersed by an \'echelle grating (P3),
cross-dispersed using a transmission grating (P4) 
and focused down using an optic (P5) on to the detector
(P6). For a detailed description of the parts see 
Section~2.\ref{sec:spec}.}
\label{fig:rhea-nir-img}
\end{figure}

As illustrated in Fig.~\ref{fig:rhea-nir-img}, the
spectrograph is composed of the \ac{MCF} with the 3D 
printed \ac{MLA} (P1), a combination of optical lenses (P2) 
to collimate the beam (EO\#49-656-12mm, EO\#47-655-INK-36mm,
Thorlabs TTL200-A) an \'echelle R2 grating (P3) (Thorlabs
GE2550-0363) for dispersion, then cross-dispersion
using a transmission grating (P4) ($2''$ 200\,l/mm Baader
Planetarium) and finally a lens (P5) (Thorlabs TTL200-B)
to focus the beam onto the detector (P6) (ASI-183MM-pro
mono).

\ac{3D-M3} makes use of only the 7 inner cores of the \ac{MCF}
out of 19 in total in order to avoid overlapping of cores
on the detector. A custom aperture to mask off the outer 
ring of 12 cores was positioned in between the \ac{MCF}
exit and the FC/PC connector mounted on the spectrograph
side. The \ac{MCF} was rotated around its central core 
until all the spectra from the individual cores were well
separated and equidistant (see Fig.~\ref{fig:spectrum}
bottom panel \new{and Ref.~\cite{Leon-Saval:2012,Betters:2014}}).
The resolving power of the spectrograph 
was measured as 30,000 at 1\,$\upmu$m, while the footprint 
of the spectrograph is $300\times600$\,mm. It was installed
on the Nasmyth \ac{NIR} platform of the Subaru Telescope.

\section{Results}
\label{sec:results}
\subsection{Laboratory throughput results}
\label{sec:lab-th-res}
To characterize the performance of the injection system 
prior to on-sky operation, we performed throughput 
measurements in lab conditions as described in Section~2.\ref{sec:throughput}.
The results are presented here.

\ac{SCExAO} is equipped with an internal calibration
system using a Fianium supercontinuum source to inject
light using an endlessly-\acf{SMF} delivering a broadband 
(from visible to K-band) diffraction-limited \ac{PSF} 
into the facility following the same optical path as the
light entering from the Subaru Telescope and the AO188
facility. This source was used to inject light into
\ac{SCExAO} (see Fig.~\ref{fig:inj-scexao}) reaching 
the 3D printed \ac{MLA} facet. Frames with exposure times
of a fraction of a second were acquired with both fibers,
the \ac{MCF} and the large \ac{MMF} used as a reference,
using the experiment apparatus described in Section~2.\ref{sec:throughput}. 
Dark frames were also recorded,
which were averaged and subtracted from the data frames 
before further processing.

Two different experiments were performed to calculate 
the throughput performance of the \ac{IFU}; in the first
the maximum coupling of light into each core was 
calculated by measuring the optimal throughput after
centering the incoming light on the fiber core, and in
the second the sensitivity of the coupling efficiency to
mis-alignment of the injected beam was calculated by
measuring the throughput of the central core of the
\ac{MCF} as the injected beam was laterally displaced by
steps of 5\,$\upmu$m \new{(9\% of the each lenslet diameter)}
in order to evaluate the performance of the fiber in a 
more realistic scenario, where the target affected by 
atmospheric perturbations will be moving.

Through simulation we determined that the \ac{MMF}
coupled all of the injected light. Therefore, we could
use the power coupled into the \ac{MMF} to normalize
the flux from the \ac{MCF}. To this end, we scanned
the 5-axis stage to optimize the coupling into each
core.

\begin{figure}
\centering
\includegraphics[width=\columnwidth]{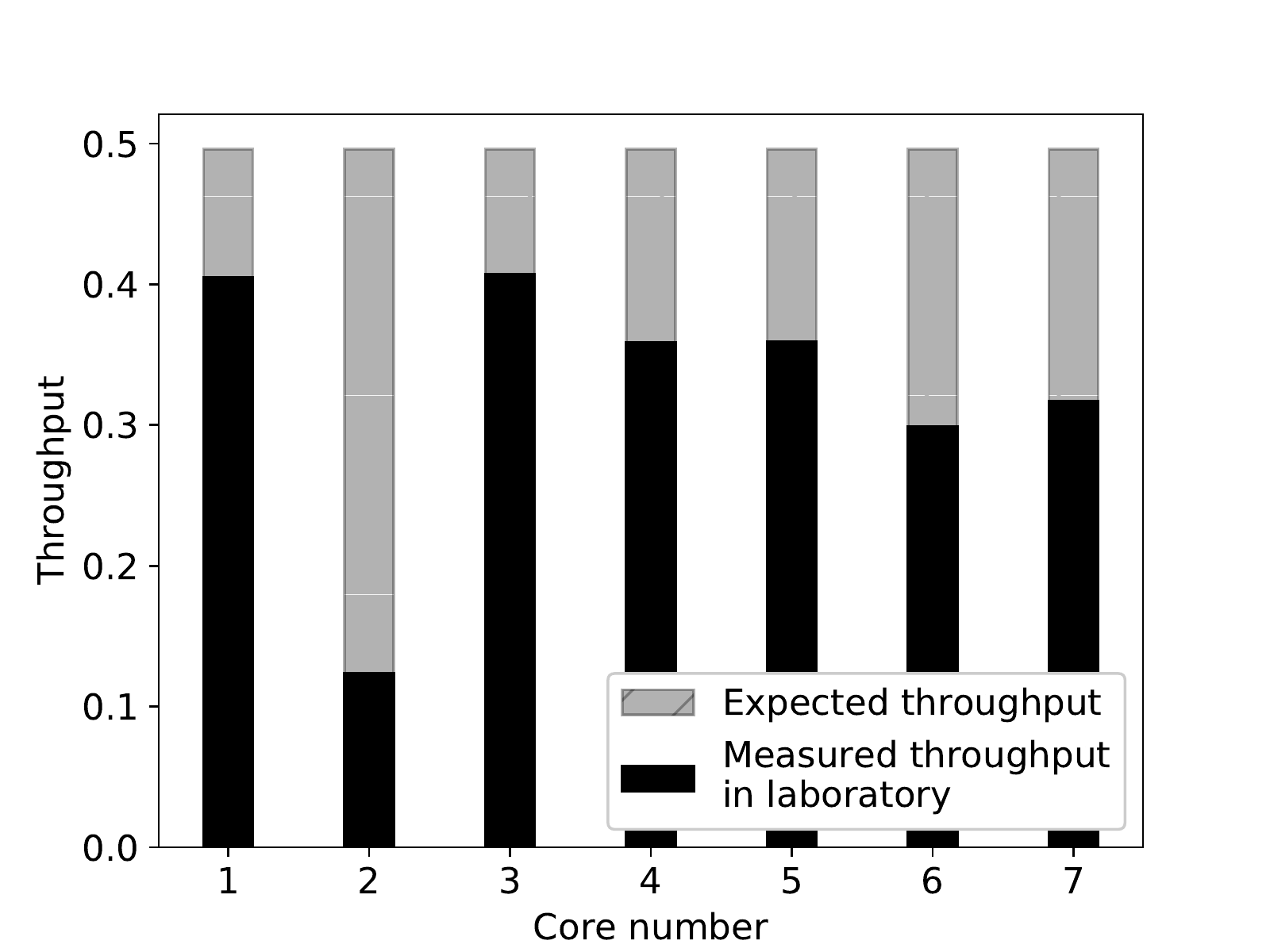}
\caption{Throughput efficiency for each of the 7 cores
of the fiber (see Fig.~\ref{fig:mcf-x-sect} for the 
correspondence of core\# to position on the fiber).
The injected $f/\#$ was fixed to $f/$57.5 by positioning the
5-axis mount and the fiber to the desired location.}
\label{fig:mcf_th_lab}
\end{figure}

The results were derived using the \ac{PIAA} optics
of \ac{SCExAO}. The outcome of the first experiment 
are shown in Fig.~\ref{fig:mcf_th_lab}. The average
throughput was 35.8\,$\pm$\,1.6\% with a maximum 
equal to 40.7\,$\pm$\,2\%. 
From the calculation of the average the low throughput of
the fiber core\,\#2 (12.3\,$\pm$\,2\%) was excluded as 
it was not representative of the rest of the lenslet structure
due to manufacturing errors. The maximum achieved throughput
corresponds to 83 per cent of the theoretically expected
value (48.8\%) according to \texttt{Zemax} calculations
\new{(see Section~2.\ref{sec:sims})}. The throughput across
all the 7 lenslets 
when illuminated by a single, unresolved star was measured
to be 70\,$\pm$\,3\% while the simulated is 83\% (84.8\%
of the theoretical value
\new{The residual throughput loss of
$\sim$8\% can be attributed to Fresnel reflections 
($\sim$4\%), impurities of the imperfectly polished fiber
facets (less than few percent estimated based on 
past lab experience), and mode-field mismatch at the
focus (3-5\%).}

\begin{figure}
\centering
\includegraphics[width=\columnwidth]{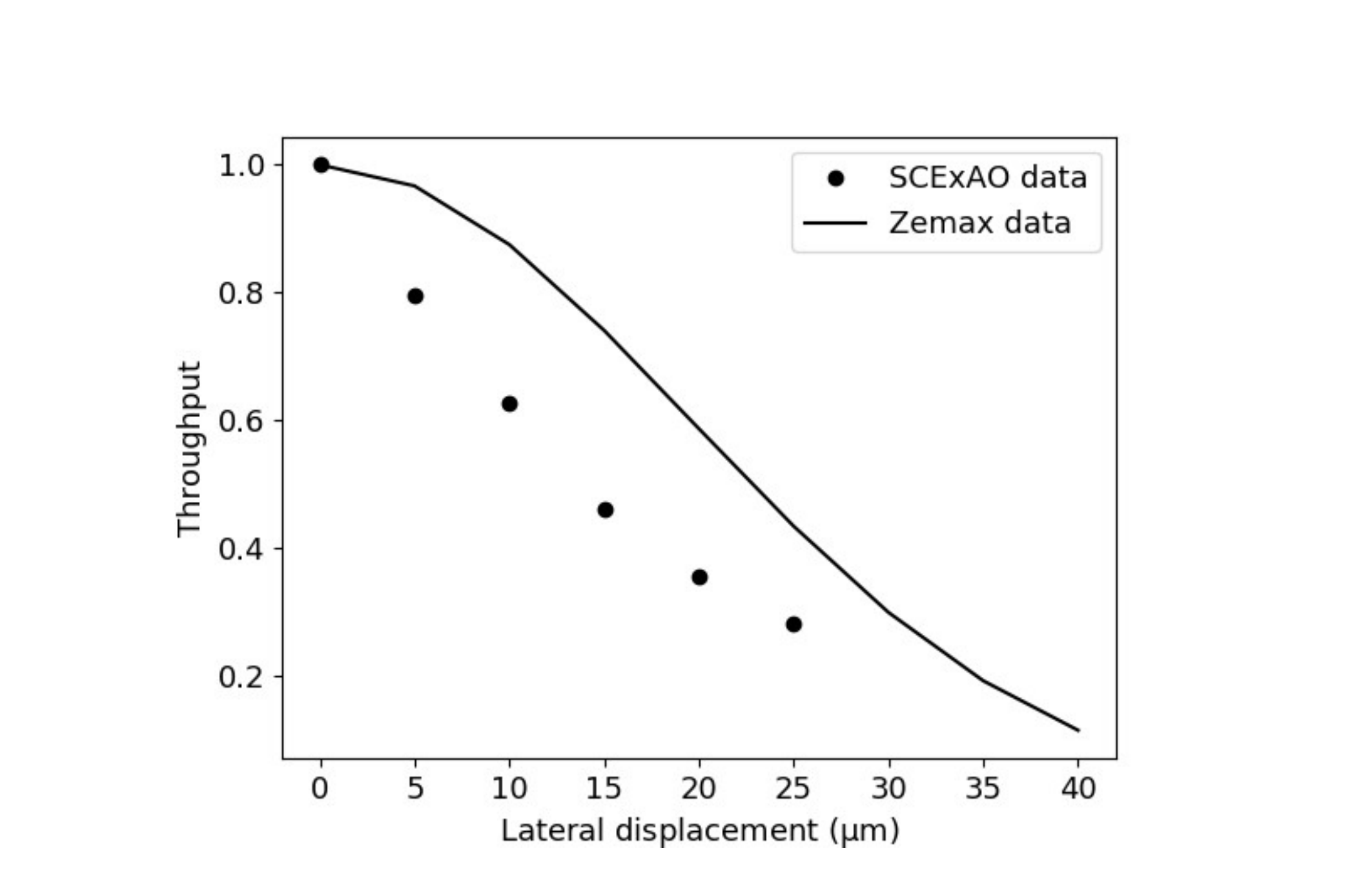}
\caption{Throughput efficiency of the central core
as a function of lateral displacement for \ac{SCExAO}
data compared to simulations with \texttt{Zemax}.
Results are normalized to the maximum throughput and
error-bars are smaller than the data points.}
\label{fig:mcf_th-dis_lab}
\end{figure}

The results of the second experiment are shown in
Fig.~\ref{fig:mcf_th-dis_lab}, which shows that 
the mis-alignment tolerances of the injected beam are
relaxed using the micro-lenses, maintaining the 
throughput \new{to within 35\% of the peak even for a
lateral $\sim$20\,$\upmu$m off-axis injection.}

\subsection{Turbulence impact on the throughput}
\label{sec:sim-turb}

Figures \ref{fig:mcf_th_lab} and \ref{fig:mcf_th-dis_lab}
represent the throughput efficiency of ideal
atmospheric conditions with a nearly perfect wavefront.
However, under realistic atmospheric conditions the
performance is reduced as a function of the injected
wavefront errors post \ac{AO} correction. In order to
assess the coupling efficiency of the starlight 
into the \ac{IFU} system the throughput across 
all of the 7 cores of the \ac{MCF} was measured,
representative of an unresolved target. The 
measurements were performed using the turbulence
simulator of \ac{SCExAO} in laboratory conditions.
A detailed description of the turbulence simulator
can be found in Ref.~\cite{Jovanovic:2017,Jovanovic:2015}.

\begin{figure}[ht]
\centering
\includegraphics[width=\columnwidth]{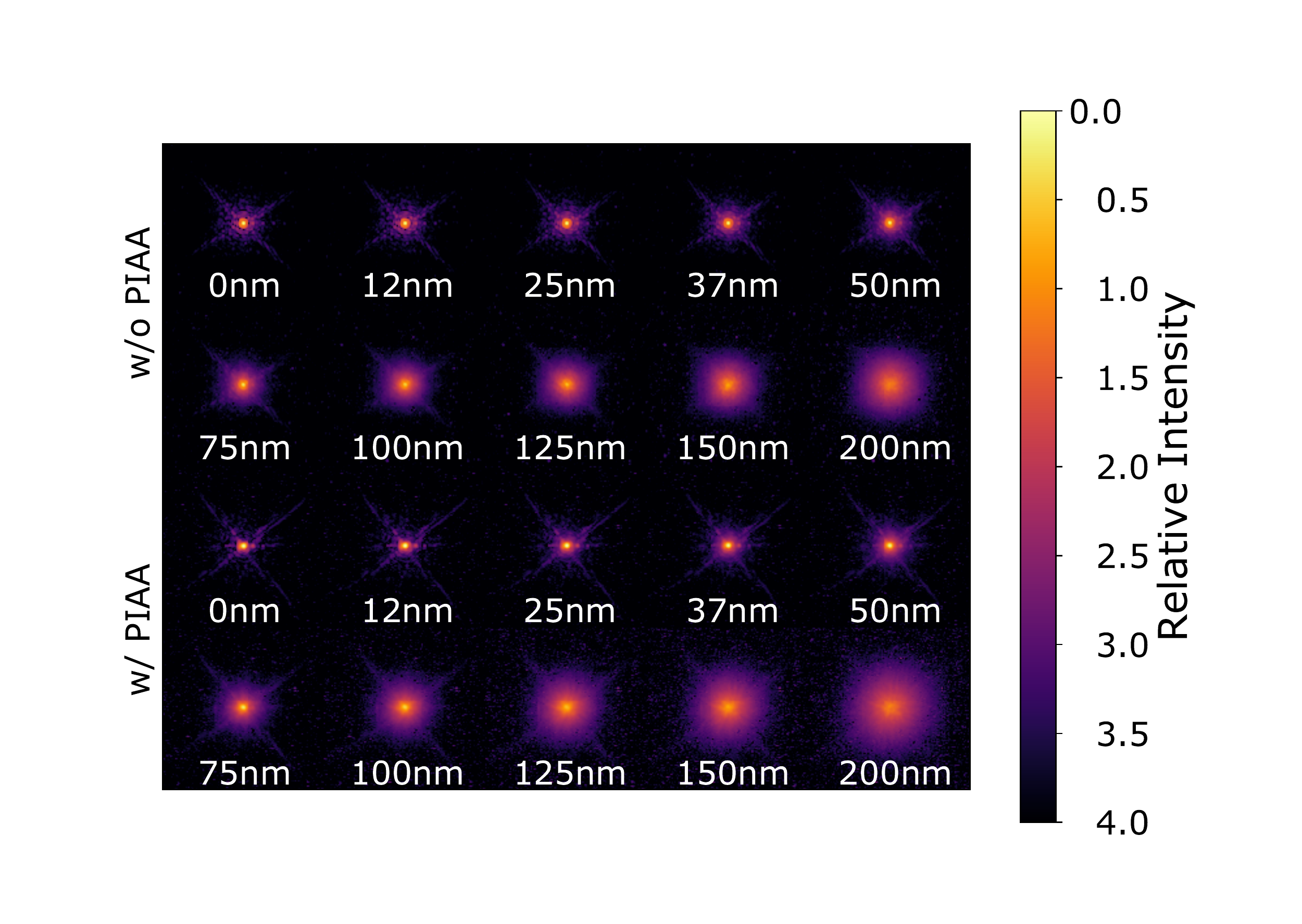}
\includegraphics[width=\columnwidth]{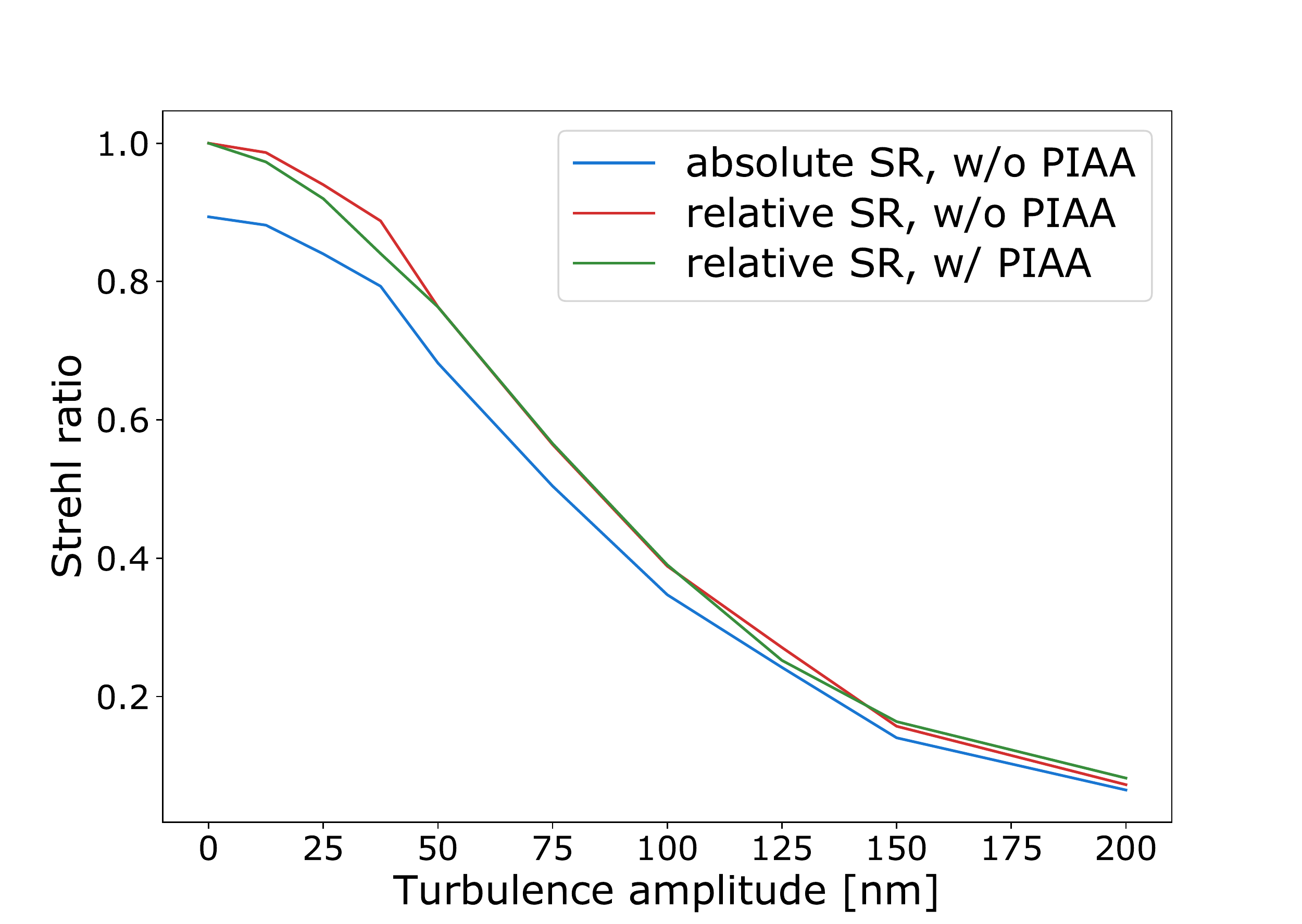}
\caption{\textbf{Top panel}: Images of the injected
beam at the focal plane of the internal camera 
at the \ac{SCExAO} bench as a function of turbulence 
amplitude. This is shown without (top to middle) and with
(middle to bottom) the \acl{PIAA} optics in the optical train.
\textbf{Bottom panel}: Calculated \acl{SR} as a function of
turbulence amplitude using the \acl{PIAA} optics (green)
and without them for the absolute (blue) and relative (red)
\acl{SR} measurements.}
\label{fig:th-sr-inj}
\end{figure}

The turbulence simulator was configured
to produce a range of atmospheric turbulence amplitudes 
to replicate a variety of on-sky conditions ranging
from 0-200\,nm RMS \new{wavefront errors} with variable steps of 12.5 and 25\,nm.
This was done to address the coupling performance 
of the \ac{IFU} under different \acp{SR}. Three types
of data frames were captured: 1)\,\ac{PSF} with the 
\ac{PIAA}, 2)\,\ac{PSF} without the \ac{PIAA} and 
3)\,dark data images for the background subtraction of
the previous two data frames. Following the data 
reduction, the \ac{SR} as a function of the 
turbulence amplitude was calculated. Using the data
frames without the \ac{PIAA} optics, the reduced data
frames were compared with a simulated y-band 
(960-1080\,nm) \ac{PSF} in order to work out the 
absolute \ac{SR} as a sanity check. The maximum 
measured \ac{SR} was found to be $\sim$90\% as a 
result of the image sampling that is close to the 
Nyquist criterion in addition to uncorrected
low-order errors left on the optical train. After
that, the measured \ac{PSF} was used as our reference 
for the \ac{SR} derivation with the \ac{PIAA}. The 
ten variations of injected \acp{PSF} as well as the 
calculated \acp{SR} as a function of the turbulence
amplitude are presented in Fig.~\ref{fig:th-sr-inj} 
as captured by the internal camera of the \ac{SCExAO}.
On the top sub-figure of Fig.~\ref{fig:th-sr-inj} from
its top to middle, the ten injected \acp{PSF} are shown
without the \ac{PIAA} optics while in the middle to
bottom frame the corresponding to those acquired with
the usage of \ac{PIAA} optics on train.

\begin{figure}
\centering
\includegraphics[width=\columnwidth]{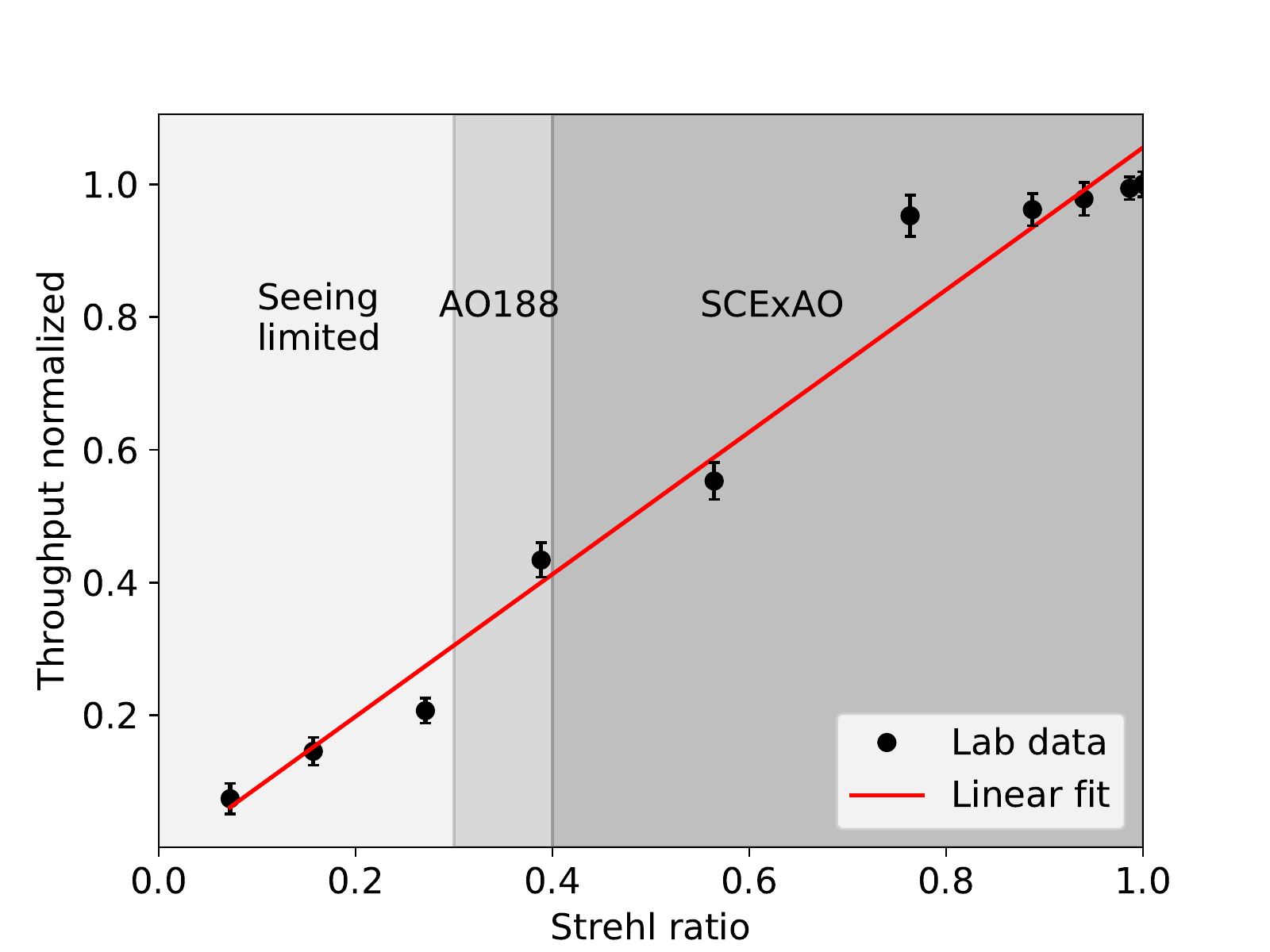}
\caption{Normalized throughput efficiency as a 
function of \acl{SR}, calculated using laboratory
data in the y-band (960-1080\,nm). A linear fit is
overplotted onto the data points and the seeing 
limited, AO188 and \ac{SCExAO} correction regimes
are shown.}
\label{fig:th-SR}
\end{figure}

The output flux of the \ac{IFU} was measured using the
above mentioned setup in Section~2.\ref{sec:throughput}.
The results from this experiment are presented in 
Fig.~\ref{fig:th-SR}. This figure shows the relation
of the normalized throughput efficiency as a function
of the measured \ac{SR}. A linear fit was performed 
to highlight the linear response of the
throughput as a function of the \ac{SR}. The mathematical
equation of the linear fit at 980\,nm is given by
$\eta = \ac{SR} \times 1.07 - 0.01 (\%)$, where $\eta$
is the coupling efficiency. This means theoretically 
that the coupling efficiency using the \ac{SCExAO} can 
be as high as 24-33\% provided that a \ac{SR} of
60-80\% is achieved for the fine platescale of 18\ac{mas}.
As the fiber was found to be damaged near the non 
3D-printed end, the following measurements are normalized.
This then neglects the damage to the fiber, whilst 
showing the performance of the \acp{MLA}.

\subsection{On-sky photometric results}
\label{sec:on-sky-res-th}

The on-sky performance of the \ac{IFU} system was
tested on the night of 7th of October 2020 during
a \ac{SCExAO} engineering run. Consequently, the
true performance of the system was measured under 
real observing conditions. However, as mentioned
above the \ac{IFU} cable was damaged and thus the
maximum expected throughput performance was not achieved.

The starlight follows the optical path as described
in Section~2.\ref{sec:scexao-infr}. Similarly to the
laboratory experiments, the \ac{PIAA} optics were
used to apodize the injected beam.

In order to evaluate the on-sky conditions, right
after the end of the observation, data of the
injected \ac{PSF} was gathered for $\sim$4\,min.
These data are from a non-common focal plane with
the \ac{IFU} (see bottom left of Fig.~\ref{fig:bench-scexao}).

The photometric image data were acquired using the 
setup described in Section~2.\ref{sec:throughput}, 
(see Fig.~\ref{fig:th_setup}).
The image data were taken with exposure times of a 
fraction of a second. Three types of on-sky data 
frames were gathered: 1)\,output data images of the 
focal plane of the \ac{IFU}, 2)\,dark data images for
the background subtraction of the output data images
and 3)\,bias frames for pixel-to-pixel variation
correction. Following that, the averaged bias and 
dark data images were subtracted from the output 
images as well as abnormal pixel values (hot, dead).

Two different targets were selected to be observed
in order to characterize the efficiency of the 
\ac{IFU} system, a target resolved by the Subaru 
Telescope and an unresolved one. The selected targets
were Mira ($o$ Ceti), as a representative of resolved
target with an angular dimension of $\sim$50 \ac{mas}
and $\delta$ Psc, an unresolved target with an angular
diameter of about $\sim$4~\ac{mas}. The atmospheric
conditions during the night of observation were within 
acceptable values (mean Seeing 0.42$''$, mean humidity
30\%, \url{http://mkwc.ifa.hawaii.edu/current/seeing/images/20201007.meteogram.jpg})
resulting in a mean \ac{SR} of 0.4 in the y-band as
calculated.

\begin{figure}
\centering
\includegraphics[width=\columnwidth]{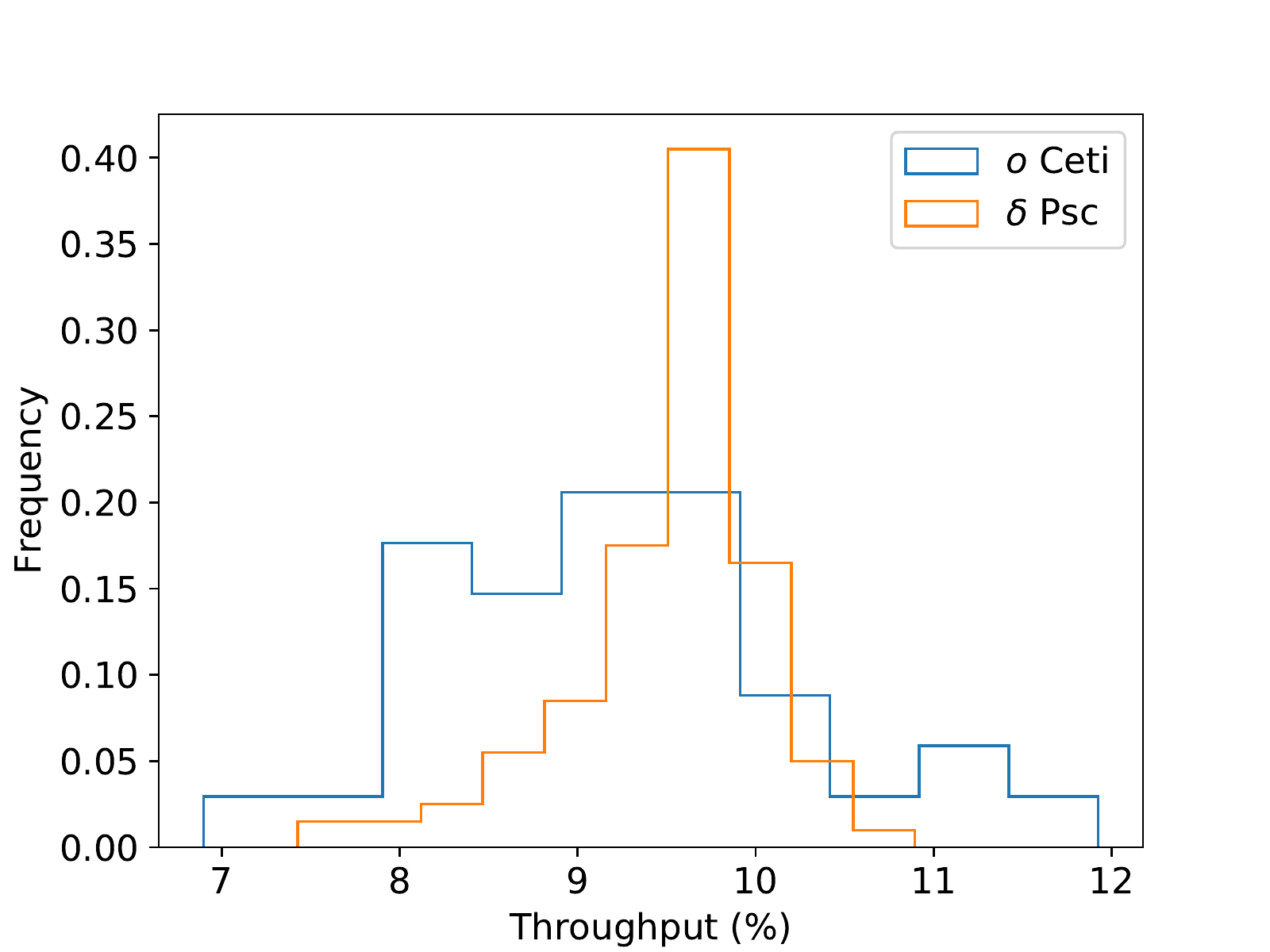}
\caption{Calculated on-sky throughput results as a function 
of the occurrence rate for $o$ Ceti and $\delta$ Psc.}
\label{fig:on-sky-th}
\end{figure}

Numerous image data sets were captured on the photometer
on both targets and the best overall throughput summed 
from all the 7 cores was 11.9\,$\pm$\,2.5\% for the
$o$ Ceti and 10.9\,$\pm$\,3\% for $\delta$ Psc. As 
mentioned in Section~3.\ref{sec:lab-th-res} the laboratory
measured throughput across all the 7 lenslets for the
case of a single un-resolved star is 70\%.
The coupling results in the experiment were limited by
the above mentioned damaged fiber. The photometric results
are presented in Fig.~\ref{fig:on-sky-th}. From the 
figure it can be seen that the dispersion around the
maximum throughput highlights the \ac{MLA} performance
in keeping the coupling uniform even with the fine
platescale of 18\,\ac{mas}. That produces uniform
spectral traces of the individual \ac{MCF} cores
on the spectrograph detector for a resolved target,
while a great portion of the light of an unresolved target 
is sampled mostly by the central core of the fiber.

\subsection{On-sky spectroscopic results}
\label{sec:on-sky-res}

\begin{figure}
\centering
\includegraphics[width=\columnwidth]{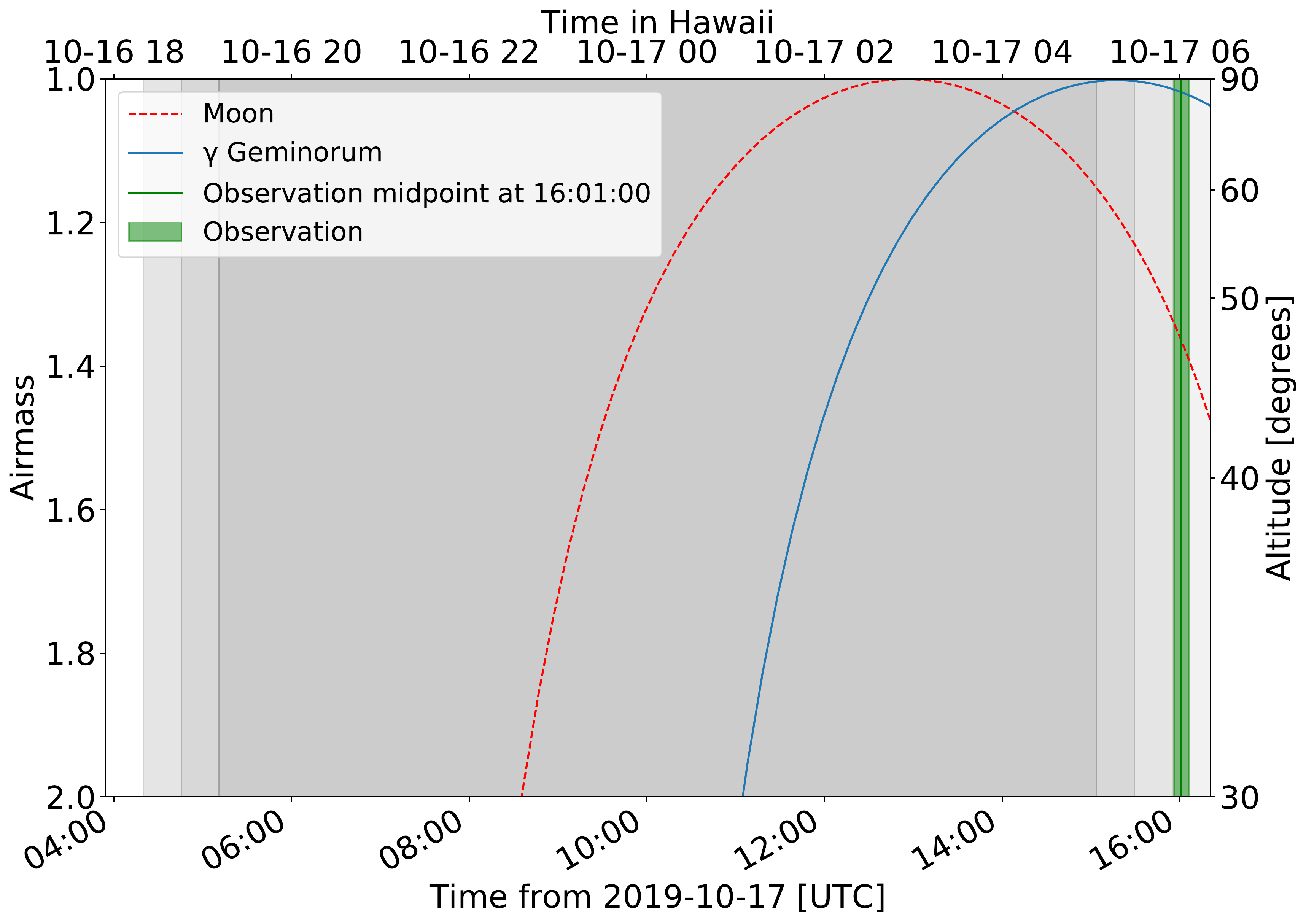}
\caption{Illustration of time vs. airmass/altitude 
of the observation at 6:01 UTC -10 on the 17th of
Oct. 2019 from Mauna Kea Observatory (green vertical
region), for $\upgamma$ Geminorum (solid blue line)
and the moon (red dashed). Exposure started at the 
very end of observing time, close to sunrise.}
\label{fig:gamma-gem-int}
\end{figure}

To evaluate the on-sky performance of the \ac{3D-M3} 
instrument, it was tested on the night of 16$^{th}$ 
of October 2019 during a \ac{SCExAO} engineering 
run. Thus, the true performance of the \ac{3D-M3} 
was measured under real observing conditions. However,
due to restrictions in available on-sky time, only
one exposure of 8\,min duration was captured at 6:00
(UTC-11) (see Fig.~\ref{fig:gamma-gem-int}).

As explained in Section \ref{sec:scexao-infr},
starlight from the Subaru Telescope enters through the 
AO188 \cite{Minowa:2010} into the \ac{SCExAO} 
instrument. With the intent of maximizing the starlight
coupling into the \ac{IFU}, the \ac{WFS} loop was
closed and the \ac{PIAA} optics deployed. Specifically, 
additional software modules were used to improve the
sensitivity by predictive control applied on the \ac{WFS}
to remove resonances \cite{Poyneer:2014} and enhance the
wavefront correction. In addition, \ac{PIAA} optics 
apodize the injected beam as was the case for the 
laboratory experiments. 

After the end of the observation, \ac{PSF} data of
the injected beam was collected for $\sim$4\,min in
order to assess the on-sky conditions. However, the 
acquired data were from a non-common focal plane
with the \ac{IFU}, so there were non-common path
errors as expected (see bottom left of Fig.~\ref{fig:bench-scexao}).

\begin{figure*}
\centering
\includegraphics[width=1.5\columnwidth]{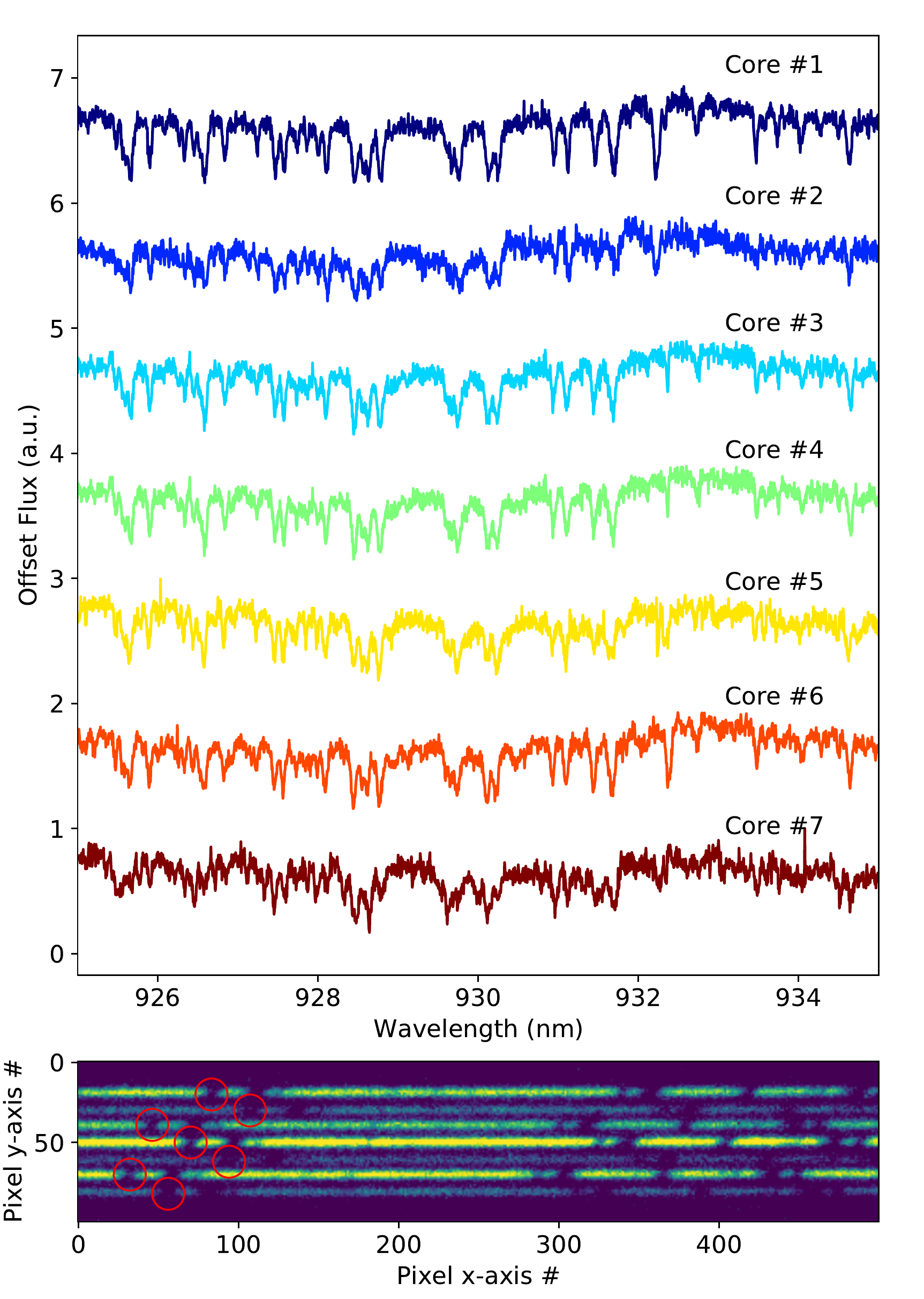}
\caption{\textbf{Top panel:} Spectra of $\upgamma$ 
Geminorum from the 7 individual cores of the
\ac{MCF} in the wavelength range of 925-935\,nm where the
many telluric absorption features are visible. Each 
spectrum is self-normalized and offset with respect to
its intensity for clarity. \textbf{Bottom panel:} 2D 
detector image of a part of a spectral order from
$\upgamma$ Geminorum. All 7 cores are visible with
sampled starlight. Notice the rotated pattern of the 
\ac{MCF} formed by absorption features marked in red 
circles. On both axes numbers represent the detector 
pixels 4.8$\times$4.8\,$\upmu$m.}
\label{fig:spectrum}
\end{figure*}

\begin{figure}
\centering
\includegraphics[width=\columnwidth]{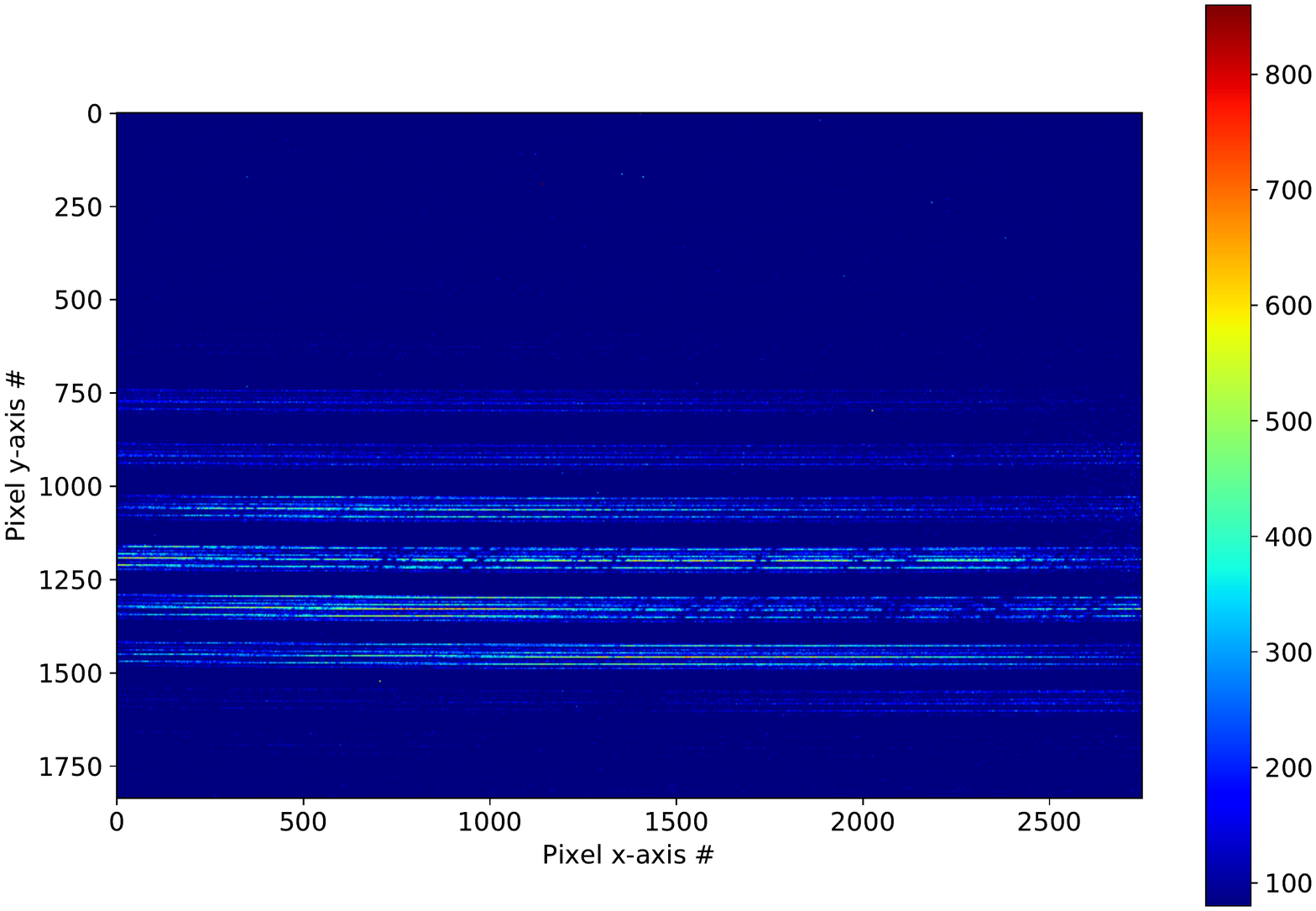}
\caption{Image of the spectrograph detector of the $\upgamma$ Geminorum.}
\label{fig:f-spec}
\end{figure}

The spectroscopic image data were acquired with the 
ASI-183MM-pro mono CMOS detector with 8\,min of
exposure time. For the spectroscopic analysis four
types of on-sky data images were collected: 1)\,output 
data images were taken at the focal plane of the 
spectrograph, 2)\,dark frames that were used for
the background subtraction of the the output data
images, 3)\,bias frames for correction of the
pixel-to-pixel variations and 4)\,flat frames 
for the output normalization. Following the 
acquisition of data images, the averaged dark and
bias frames were subtracted from the output images,
abnormal (hot, dead) pixels were removed, and the
averaged flat frames were used to normalize the final
spectrum image data. The reduced frame is presented 
in Fig.~\ref{fig:f-spec}.

The target was $\upgamma$ Geminorum, which is an 
A1.5IV+ sub-giant star expected to be almost free 
of absorption features, except for broad hydrogen 
lines. The resulting low \ac{SNR} spectrum 
(SNR$\sim$9) from this observation is shown in 
Fig.~\ref{fig:spectrum}. In this figure, only the
spectrum over the range of 925-935\,nm is presented 
(908-968\,nm is the full range), focusing on a narrow 
region where the spectral absorption features that
are caused by the Earth's atmosphere are most intense.
The figure shows the spectra for all of the 7 cores
individually. In the bottom panel of the figure, the 
2D image of a part of the full spectrum is included 
for visual clarity of the geometry. The hexagonal
pattern of the \ac{MCF} at optimal rotation appears 
as 7 lines for each order.

\begin{figure*}
\centering
\includegraphics[width=0.9\textwidth]{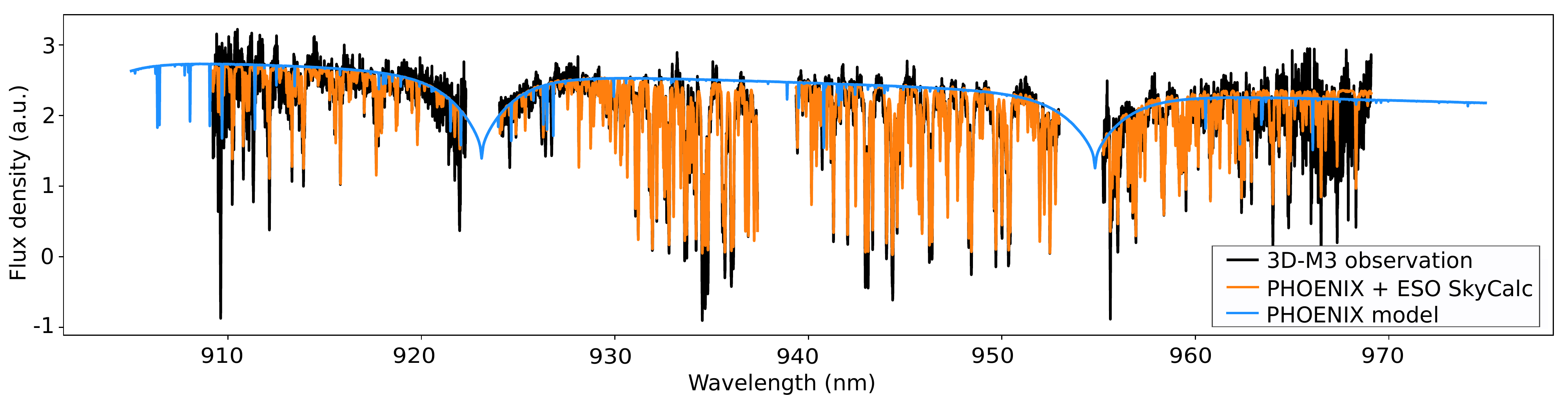}
\caption{Full spectrum of $\upgamma$ Geminorum derived 
from combining all the fiber cores overplotted (in blue)
with a PHOENIX synthetic star spectrum \cite{Husser:2013},
and also (in orange) including telluric absorption estimated
with ESO's SkyCalc \cite{Noll:2012,Jones:2013}. The model 
spectra were convolved down to R$\sim$30,000 for the 
comparison. The two broad features in the stellar spectrum
are the Paschen 8 and 9 lines.}
\label{fig:f-spec-comp}
\end{figure*}

In the next step the spectra from each of the 7 cores
were summed together in order to further increase the 
\ac{SNR}, representative of the case of a single
unresolved star spectrum observation. The resultant
combined spectrum is presented in Fig.~\ref{fig:f-spec-comp}.
The observed spectrum is represented with black color, 
with blue color the model spectrum for an $\upgamma$
Geminorum-like star from the PHOENIX spectral library 
\cite{Husser:2013} without the effect of telluric 
absorption, and with orange color the PHOENIX model 
spectrum with telluric absorption as calculated using
ESO's SkyCalc \cite{Noll:2012,Jones:2013}.

\subsection{Adaptive optics performance}
\label{sec:ao-performance}
To quantify and evaluate the performance of the \ac{AO}
correction, the on-sky \ac{PSF} profile data were 
captured as explained in Section~3.\ref{sec:on-sky-res}. 
The \ac{PSF} was recorded in \ac{FITS} format, with data cubes
composed of 1000 instantaneous frames each, where the
total exposure time of each data cube was $\sim$0.83\,sec
in total. For calibration purposes, additional \ac{PSF}
data cubes were captured off-sky as well using the 
internal calibration source of \ac{SCExAO}, which 
follows the same optical path as the starlight (see 
center left of Fig.~\ref{fig:bench-scexao}) and 
represents the case of an ideally shaped and stabilized
target. Lastly, dark frames were collected for background
subtraction.

Initially, each of the 1000 frames in the 
data cubes, for off and on-sky data, were dark
current subtracted and averaged. After this, the
ratio of the maximum intensity ($I\mathrm{max}$) 
to 90\% of corresponding flux ($Flux$) was calculated
and stored for all of the data. To approximate the 
Strehl ratio during the observation time, the stored
maximum to flux ratios for each data-set (off and 
on-sky) were divided; this results in

\begin{equation}
{S = \frac{I\mathrm{max_{on-sky}}}{Flux_\mathrm{on-sky}}}
{\frac{Flux_\mathrm{off-sky}}{I\mathrm{max_{off-sky}}}}.
\label{eq:strehl}
\end{equation}

The result of the calculation was $S$\,=\,0.2, representing
an estimation of the Strehl ratio on-sky. \new{From this analysis
we confirm the visually observed low Strehl for the
spectroscopic observation and highlight the reduced Strehl
with respect to the night when the photometry was collected.}

\begin{figure}[ht]
\centering
\includegraphics[width=\columnwidth]{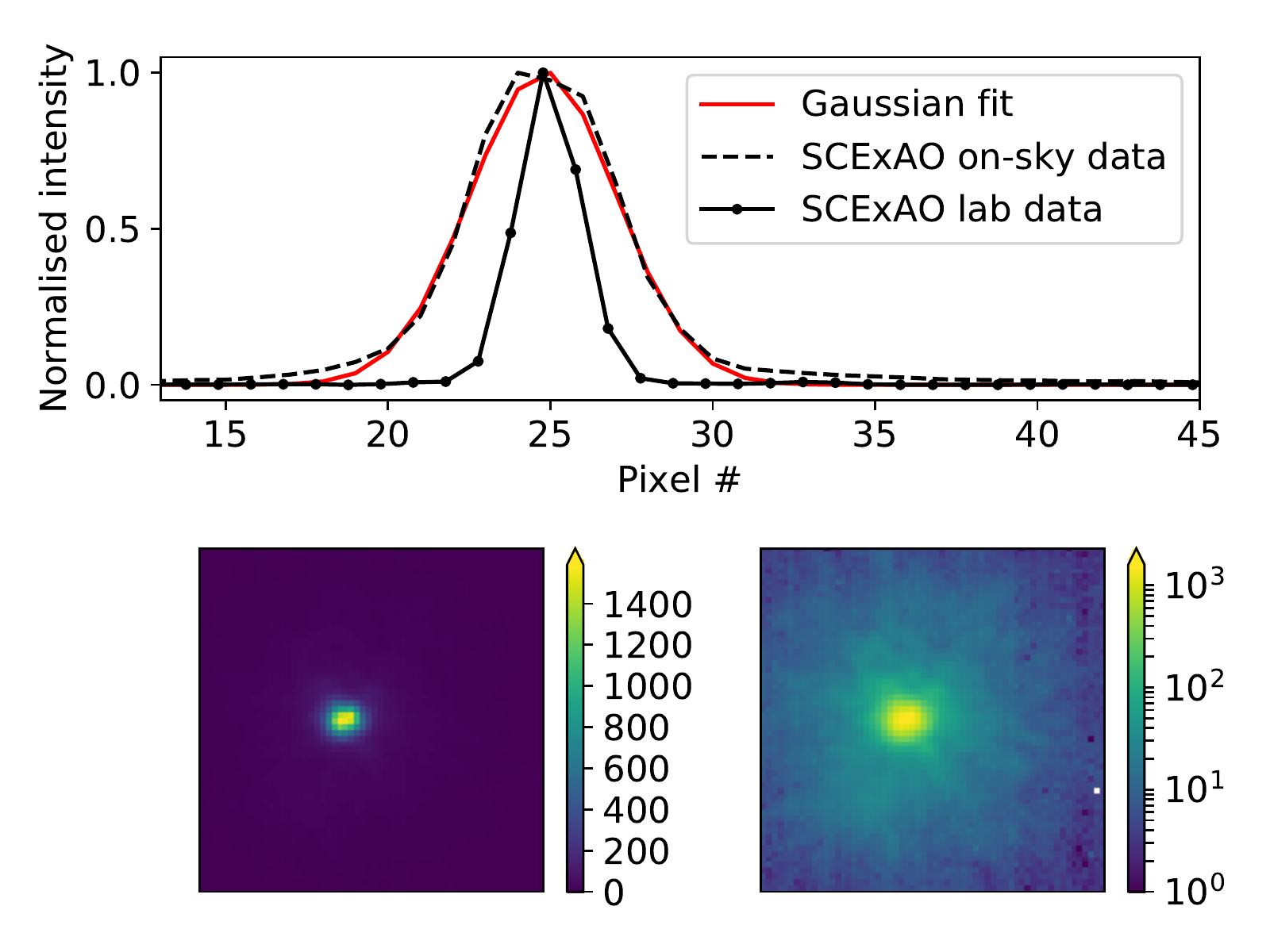}
\caption{\textbf{Top panel:} Intensity profile of
the apodized injected \acl{PSF} normalized to its peak
using the \acl{PIAA} optics, for the on-sky and
laboratory data. A Gaussian fit was performed to the
on-sky data. \textbf{Bottom panel:} 2D image of the
\ac{PSF} on-sky in linear (left) and logarithmic
(right) color scale for better clarity. The exposure
time was $\sim$0.83\,sec.}
\label{fig:gamma-gem-psf}
\end{figure}

In the next step, the averaged intensity of a data cube was 
plotted along with a Gaussian fit. This was done
for the laboratory \ac{PSF} data as well. The resultant
profile and fit are presented in Fig.~\ref{fig:gamma-gem-psf}.
In the top panel of the figure the normalized intensity 
profile of the averaged on-sky data along one axis, its 
Gaussian fit and the profile from the laboratory ideal 
beam measurements are plotted together. In the bottom 
panels the 2D image of the on-sky \ac{PSF} is plotted 
in linear and logarithmic scale for better clarity.

\begin{figure}
\centering
\includegraphics[width=\columnwidth]{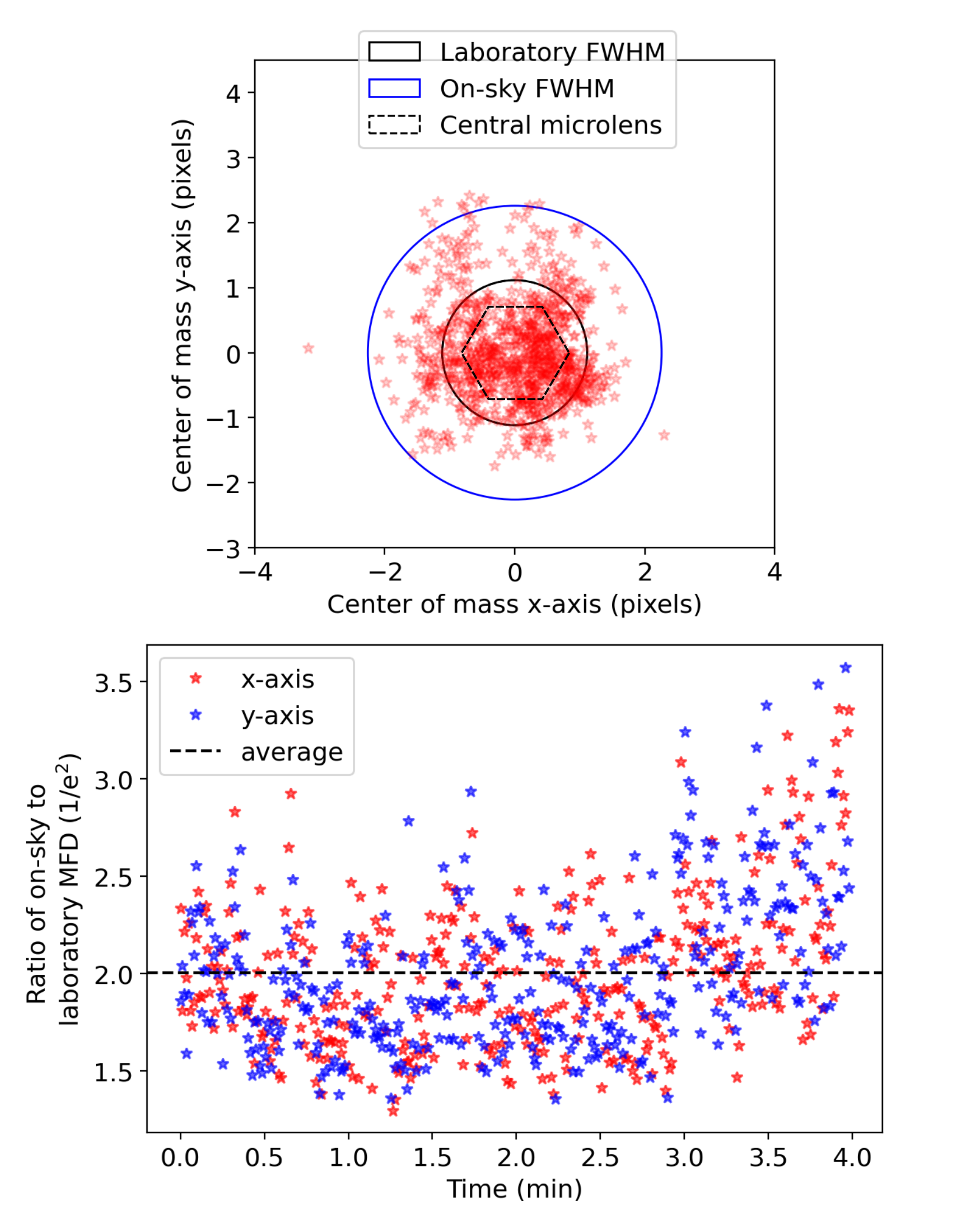}
\caption{\textbf{Top panel:} Center of mass shift of
the on-sky beam during the observation. Over-plotted with
two circles are the \ac{FWHM} of the on-sky beam and the 
laboratory beam while the dashed hexagon is the effective
area of the central micro-lens. \textbf{Bottom panel:}
The ratio of on-sky to the laboratory \ac{FWHM} for both 
detector axes for the same time duration.}
\label{fig:psf-move}
\end{figure}

Following that, the 1000 data images from each data cube
of the on-sky data, were averaged per data cube. Afterwards,
the center of mass for each of them was calculated. In
addition, the ratio of the on-sky to the laboratory
beam \ac{FWHM} was calculated as a function of time 
through the exposure to visualize the beam 
diameter changes as experienced on-sky. Results are 
presented in Fig.~\ref{fig:psf-move}; they suggest 
that the center of mass shift of the on-sky beam was 
rather big. Furthermore the size of the on-sky beam
was twice as large compared to the laboratory one as 
illustrated in the bottom panel of the figure. Consequently, the 
position stability of the \ac{PSF} was not representative
of the typical \ac{SCExAO} system performance but it was a 
consequence of the atmospheric conditions mentioned in 
Section~3.\ref{sec:on-sky-res} and the time of the observation.

\section{Discussion}
\label{sec:discussion}

\subsection{Laboratory throughput limitations}
One of the crucial benefits of the 3D-printed \acp{MLA}
is to significantly reduce the off-axis mis-alignments 
of the injected beam to a \ac{SMF}, and as a consequence
the light loss during the free-space coupling is reduced.

The results of the intended lateral beam displacements 
as shown in Fig.~\ref{fig:mcf_th-dis_lab} suggest that 
there is a difference between the expected and the 
laboratory measured performance. The difference ranges
between 0 and $\sim$27\% depending on the position
of the injected beam. This is caused by several factors
sorting them by their importance are, the high sensitivity
on the the \ac{PIAA} lenses alignment that apodize the
beam, the surface roughness of the \ac{MLA} structure
that is limited by the manufacturing process at the moment
and any mismatch of the injected
$f/\#$.

\new{The actions to reduce the influence of these
factors involve, the more accurate alignment of the 
\ac{PIAA} lenses regularly to confirm the position 
as well as the shape of the apodized beam.
An additional step is the validation
of the injected $f/\#$ of the beam by imaging it
before its entrance in the \ac{IFU}. Moreover, a 
better physical protection layer of the fiber will be
implemented to avoid any future damaged to the fiber
when it is exposed in the telescope environment.}

\subsection{On-sky observation limitations}
\label{sec:on-sky-disc}
As explained in Sections~3.\ref{sec:on-sky-res} and
3.\ref{sec:ao-performance}, the atmospheric conditions 
influenced the observation (mean Seeing 0.62$''$, high
humidity over 70\%, 
\url{http://mkwc.ifa.hawaii.edu/current/seeing/?night=20191017&lastURLnight=20191017}).
Additionally, at the time of 
the observation (sunrise) the sky brightness was not 
negligible and thus it was driving the \ac{WFS} close to 
its saturation levels, leading to an under-performing \ac{AO}
system. Furthermore, the wind was rather high in strength
and variable in direction (16\,km/h), inducing vibrations
to the structure of the telescope. Consequently, the
stabilization and the shape of the beam coupled into 
the \ac{MLA} facet was far from ideal as shown in 
Fig.~\ref{fig:psf-move} \cite{Lozi:2018}. All the factors
mentioned above are represented in the data. In more 
detail, the target under observation, which is unresolved
by the Subaru Telescope, and the majority of its light
should have been sampled by the central core of the 
\ac{MCF} as inferred from Section~2.\ref{sec:sims}. 
However, as appears in Fig.~\ref{fig:f-spec}, the 
illumination of the fiber cores is almost uniform, 
resulting in a loss of spatial information of the 
input. A likely explanation of this is the very poor 
\ac{AO} performance in correcting the shape and the 
stabilization of the injected beam. On the other hand, 
this demonstrates that our \ac{IFU} can couple light 
from a point source rather efficiently into an \ac{SM}
spectrograph, even under poor conditions.

The expected performance of the \ac{IFU} system 
encountered on-sky can be reproduced by \texttt{Zemax} 
simulations. In more detail, the \ac{POP} module was 
used to simulate the performance of one lenslet out
of the 7, as the lenslets are considered homogeneous.
Two simulations were performed to calculate the expected 
flux output at the end facet of the \ac{MCF}; in the 
first one, representative of the ideal on-sky conditions,
we used a Gaussian beam with a 72\,$\upmu$m \ac{MFD},
which is a good approximation of the output beam profile
of the Subaru Telescope, in combination with the
\ac{SCExAO} and \ac{PIAA} optics, as described in 
Section~2.\ref{sec:sim-arch}. In the second one, we used a 2$\times$ 
bigger beam (144\,$\upmu$m) in \ac{MFD} size, as the 
input to our simulations (as observed in Section~3.\ref{sec:ao-performance}), 
which had also a lateral
displacement term, relative to the central on-axis
fiber core equal to 11\,$\upmu$m (as observed in 
Fig.~\ref{fig:psf-move}).

\begin{figure}
\centering
\includegraphics[width=\columnwidth]{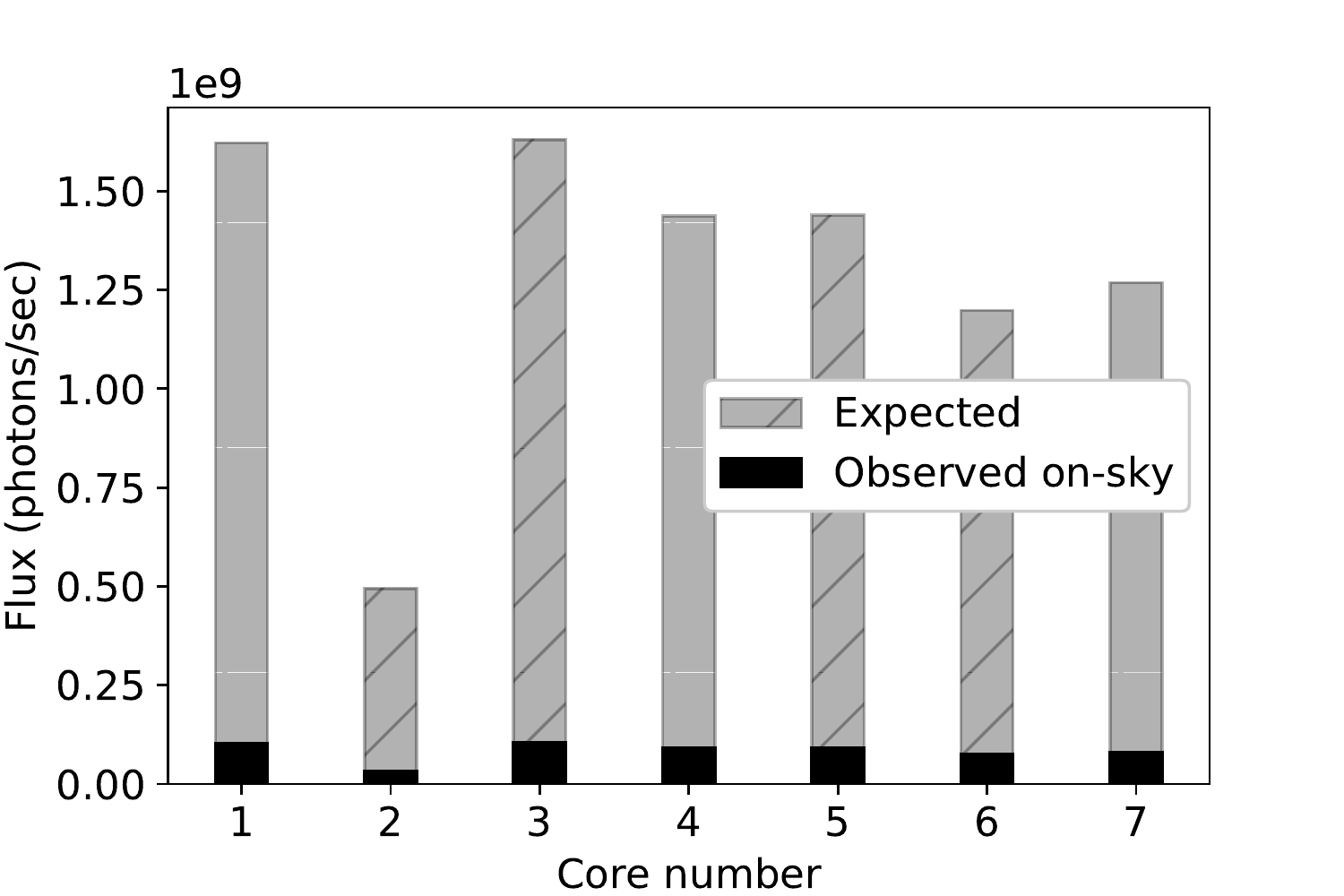}
\caption{Simulated performance of the system from
the star down to the \ac{IFU} level, for ideal and
the approximated on-sky conditions. Despite the 
relative low performance, the \ac{IFU} manifested
decent coupling from a point source into
an \ac{SM} spectrograph, even under poor conditions.}
\label{fig:flux-cores}
\end{figure}

The apparent magnitude of $\upgamma$ Geminorum is 
taken from the literature and is approximately equal 
to 1.88 in \ac{NIR} wavelengths. The transmission
properties of the sky at Maunakea as well as the 
performance of all of the optical components of the
telescope down to the injection fiber (without the 
spectrograph), for the corresponding optical band, 
were adapted from the literature and combined. The
throughput of the \ac{IFU} system was taken from 
Section~3.\ref{sec:lab-th-res}. The resulting maximum
flux derived from the simulations for the first 
experiment was found to be $1.63\times10^{9}$ 
photons/sec in the Y photometric band (970-1007\,nm)
where a bandwidth of 104\,nm of the spectroscopic
band were used. The resulting throughput for the
second experiment was found to be equal to 2.4\% 
corresponding to an average flux of $1.04\times10^{8}$
photons/sec. This is shown in Fig.~\ref{fig:flux-cores},
where the flux performance of the system up to 
the \ac{IFU} level corresponds to 6\% of the value
measured in laboratory.

The analysis above provides a solid explanation of the 
observed on-sky performance. Moreover,
the on-sky results are useful for predicting the expected
performance of the system for future reference.
To extrapolate the acquired results, given a photometric
sky and median seeing conditions when Strehl ratios are 
over 60\%, the required exposure time for achieving 
similar levels of \ac{SNR} as experienced on-sky, would
be 48\,secs instead of 8\,min.

\section{Conclusions}
\label{sec:conclusions}
In this study we accomplished the first technology 
demonstration of a high resolution diffraction-limited
integral field spectrograph, \ac{3D-M3}. It utilizes 
a custom \ac{MCF} with 3D printed micro-optics on top
of its cores in order to increase the coupling
efficiency of starlight from \ac{SCExAO} at the 8.2\,m
Subaru Telescope while the the spatial
information of the input is retained.

The injection \ac{IFU} system is optimized for an
on-sky angular dimension of 54\,\ac{mas}, using the 
output beam profile of \ac{SCExAO} that makes use 
of \ac{PIAA} optics shaping the output \ac{PSF} to
a near Gaussian profile.

The \ac{MLA} was directly 3D printed onto the \ac{MCF}
using two-photon lithography, considerably improving 
the coupling of light to each of the fiber cores 
from a few per cent levels up to a theoretical maximum of
48.8\% with the \ac{MLA}. In addition to this, the
mis-alignment tolerances of the injected beam were 
relaxed with this method, performing decently 
even for a lateral $\sim$20\,$\upmu$m off-axis injection.

The throughput results on \ac{SCExAO} confirm the
theoretical expectation. A maximum of 40.7\,$\pm$\,2\% of 
throughput per core was achieved (corresponding to
$\sim$83\% of the theoretical value) reaching an average
of 35.8\,$\pm$\,1.6\% per core. The throughput performance
of the \ac{IFU} system across all the 7 lenslets,
representing the case of a single unresolved target was
measured to be 70\,$\pm$\,3\% (corresponding to $\sim$84.8\%
of the theoretical value). The difference between the
theoretical and measured values can be attributed to 
the Fresnel losses at the interfaces of the \ac{MLA} 
with the \ac{MCF} and to their surface roughness. These 
results were obtained with \ac{SCExAO} and an artificial
light source, which closely represents the performance
of the facility under good observing conditions. A
short on-sky observation suffered from windshake and 
bad seeing, resulting in considerably lower throughout
of less than 2\% compared to the laboratory measurements
during the spectrum acquisition.

The on-sky throughput results on \ac{SCExAO} did 
not reach the expected performance as measured in 
the laboratory due to technical problems. However,
demonstrate the principle and will serve as a baseline
for future reference. A maximum throughput of 
10.9\,$\pm$\,3\% was achieved for the unresolved 
target while 11.9\,$\pm$\,2.5\% was measured for 
the case of the resolved target.

Our system met the expectations and
performed reasonably well given the conditions; the
results highlight the importance of good on-sky 
conditions allowing optimal performance. If normal
photometric conditions are met, the required exposure 
time for an observation with an SNR of 20 is reduced
by a factor of ten. Compared to conventional high 
resolution \ac{IFS} \cite{Eisenhauer:2003}, the
achieved \ac{SNR} is still low due to the performance
limitations of the optical components, mostly the 
detector efficiency in this wavelength regime.

Aims for future work include further optimization of
the optical design by performing more simulations
using \texttt{Zemax}, replacing the spectrometer
optics with custom built ones using appropriate
coatings for the wavelength range used, and investing
in the stabilization of the instrument with enclosures
and controls.

Observation aims for the future involve a list
with a variety of targets (resolvable star, 
non-resolvable, double star system, massive confirmed 
exoplanets, spectroscopic standard stars) for 
evaluating its scientific potential.

\section{Backmatter}

\begin{backmatter}

\bmsection{Acknowledgments} T.A. is a fellow of
the International Max Planck Research School for 
Astronomy and Cosmic Physics at the University
of Heidelberg (IMPRS-HD) and is supported by the 
Cotutelle International Macquarie University
Research Excellence Scholarship. M.T., M.B.,
Y.X. and C.K. are supported by Bundesministerium f\"{u}r 
Bildung und Forschung (BMBF), joint project PRIMA
(13N14630), the Helmholtz International Research 
School for Teratronics (HIRST), Deutsche 
Forschungsgemeinschaft (DFG, German Research Foundation)
under Germany's Excellence Strategy via the Excellence 
Cluster 3D Matter Made to Order (EXC2082/1-390761711).
R.J.H., P.H. and A.Q. are supported by the Deutsche
Forschungsgemeinschaft (DFG) through project 326946494,
'Novel Astronomical Instrumentation through photonic
Reformatting'.
T.B. \& S.Y. are supported from the European Union's
Horizon 2020 grant 730890, and from the UK Science and
Technology Facilities Council grant ST/N000544/1.
S.Y.H. is supported by the NASA Hubble Fellowship grant
\#HST-HF2-51436.001-A awarded by the Space Telescope
Science Institute, which is operated by the Association 
of Universities for Research in Astronomy, Incorporated,
under NASA contract NAS5-26555.
The development of SCExAO was supported by the JSPS 
(Grant-in-Aid for Research \#23340051, \#26220704 
\#23103002), the Astrobiology Center (ABC) of the 
National Institutes of Natural Sciences, Japan, the
Mt Cuba Foundation and the directors contingency
fund at Subaru Telescope, and the OptoFab node of 
the Australian National Fabrication Facility. The
authors wish to recognize and acknowledge the very 
significant cultural role and reverence that the
summit of Mauna Kea has always had within the
indigenous Hawaiian community. We are most fortunate
to have the opportunity to conduct observations from
this mountain. This research made use of Astropy, a
community-developed core \texttt{Python} package for
Astronomy \cite{astropy:2013, astropy:2018}, Numpy
\cite{numpy} and Matplotlib \cite{matplotlib}.

\bmsection{Disclosures} The authors declare no conflicts 
of interest.

\bmsection{Data Availability Statement} Data underlying 
the results presented in this paper are not publicly 
available at this time but may be obtained from the 
authors upon reasonable request.

\end{backmatter}

\bibliography{references}

\end{document}